\documentclass[a4paper,10pt]{article}
\usepackage{graphicx}
\usepackage{amsmath}
\usepackage{citesort}
\usepackage{amsfonts}
\usepackage{amssymb}
\usepackage{amsmath}
\usepackage{latexsym}
\usepackage{color}
\usepackage{ntheorem}
\usepackage{braket}

\newcommand{\myregular}{x}

\newcommand{\mnotex}[1]
{\protect{\stepcounter{mnotecount}}$^{\mbox{\footnotesize
$
\bullet$\themnotecount}}$ \marginpar{
\raggedright\tiny\em
$\!\!\!\!\!\!\,\bullet$\themnotecount: #1} }

\newcommand{\jamesx}[1]{}
\renewcommand{\jamesx}[1]{{\mnote{{\color{black}{\bf jg:}
#1} }}}

\newcommand{\h}[2]{#1\dotfill\ #2\\\ptc{fixme}}

\newcommand{\ol}{\overline}

\def\nz{\ifmmode {I\hskip -3pt N} \else {\hbox {$I\hskip -3pt N$}}\fi}
\def\zz{\ifmmode {Z\hskip -4.8pt Z} \else
       {\hbox {$Z\hskip -4.8pt Z$}}\fi}
\def\qz{\ifmmode {Q\hskip -5.0pt\vrule height6.0pt depth 0pt
       \hskip 6pt} \else {\hbox
       {$Q\hskip -5.0pt\vrule height6.0pt depth 0pt\hskip 6pt$}}\fi}
\def\rz{\ifmmode {I\hskip -3pt R} \else {\hbox {$I\hskip -3pt R$}}\fi}
\def\cz{\ifmmode {C\hskip -4.8pt\vrule height5.8pt\hskip 6.3pt} \else
       {\hbox {$C\hskip -4.8pt\vrule height5.8pt\hskip 6.3pt$}}\fi}
\def\au{{\setbox0=\hbox{\lower1.36775ex\hbox{''}\kern-.05em}\dp0=.36775ex\hs
kip0pt\box0}}
\def\ao{{}\kern-.10em\hbox{``}}

\newcommand\Gregbeq{\begin{eqnarray}}
\newcommand\Gregeeq{\end{eqnarray}}

\newcommand{\tC}{\tilde C}

\def\h1{{\hat 1}}
\def\h2{{\hat 2}}

\def\3f{\frac{3}{2}}

\newcommand{\roscoff}[1]{}

{\catcode `\@=11 \global\let\AddToReset=\@addtoreset}
\AddToReset{figure}{section}

\DeclareFontFamily{OT1}{rsfs}{}
\DeclareFontShape{OT1}{rsfs}{m}{n}{ <-7> rsfs5 <7-10> rsfs7 <10-> rsfs10}{}
\DeclareMathAlphabet{\mycal}{OT1}{rsfs}{m}{n}

{\catcode `\@=11 \global\let\AddToReset=\@addtoreset}
\AddToReset{equation}{section}

\newcounter{mnotecount}[section]

\renewcommand{\themnotecount}{\thesection.\arabic{mnotecount}}

\newcommand{\jlcax}[1]{}
\newcommand{\eean}{\nonumber\end{eqnarray}}

\newcommand{\kk}[1]{}

\newcommand{\regular}{$I^+$--regular}

\newcommand{\beq}{\begin{equation}}

\newcommand{\FS}       
                  {F}

\newcommand{\HS} 
       {H_{\mbox{\scriptsize volume}}}

{\ptc{this should be removed in the oberwolfach version}}%

\newcommand{\eeal}[1]{\label{#1}\end{eqnarray}}
\newcommand{\bed}{\begin{deqarr}}
\newcommand{\eed}{\end{deqarr}}
\newcommand{\bedl}[1]{\begin{deqarr}\label{#1}}
\newcommand{\eedl}[2]{\arrlabel{#1}\label{#2}\end{deqarr}}

\newcommand{\mcO}{{\mycal O}}
\newcommand{\mcU}{{\mycal U}}

\newcommand{\bel}[1]{\begin{equation}\label{#1}}
\newcommand{\bea}{\begin{eqnarray}}
\newcommand{\bean}{\begin{eqnarray}\nonumber}
\newcommand{\beal}[1]{\begin{eqnarray}\label{#1}}
\newcommand{\eea}{\end{eqnarray}}

\newcommand{\qed}{\hfill $\Box$ \medskip}
\newcommand{\proof}{\noindent {\sc Proof:\ }}
\newcommand{\be}{\begin{equation}}
\newcommand{\eeq}{\end{equation}}
\newcommand{\ee}{\end{equation}}
\newcommand{\beqa}{\begin{eqnarray}}
\newcommand{\eeqa}{\end{eqnarray}}
\newcommand{\beqan}{\begin{eqnarray*}}
\newcommand{\eeqan}{\end{eqnarray*}}
\newcommand{\ba}{\begin{array}}
\newcommand{\ea}{\end{array}}

\newcommand{\mnote}[1]
{\protect{\stepcounter{mnotecount}}$^{\mbox{\footnotesize
$
\bullet$\themnotecount}}$ \marginpar{
\raggedright\tiny\em
$\!\!\!\!\!\!\,\bullet$\themnotecount: #1} }

\newcommand{\warn}[1]
{\protect{\stepcounter{mnotecount}}$^{\mbox{\footnotesize
$
\bullet$\themnotecount}}$ \marginpar{
\raggedright\tiny\em
$\!\!\!\!\!\!\,\bullet$\themnotecount: {\bf Warning:} #1} }

\newcommand{\R}{\mathbb R}

\newcommand{\N}{\mathbb N}

\newcommand{\eq}[1]{(\ref{#1})}

\newcommand{\ptc}[1]{\mnote{{\bf ptc:}#1}}

\newcommand{\beqar}{\begin{deqarr}}
\newcommand{\eeqar}{\end{deqarr}}

\newcommand{\beaa}{\begin{eqnarray*}}
\newcommand{\eeaa}{\end{eqnarray*}}

\newcommand{\tr}{\mbox{tr}}

\newcommand{\hg}{{\hat g}}

\newcommand{\bethm}{\begin{Theorem}}
\newcommand{\et}{\end{Theorem}}
\newcommand{\bl}{\begin{Lemma}}
\newcommand{\What}{{\mathring W}}
\newcommand{\timW}{{\hat W}}

\newtheorem{theorem}{{\sc  Theorem}\rm}[section]

\newtheorem{corollary}[theorem]{{\sc Corollary}\rm}

\newtheorem{lemma}[theorem]{{\sc Lemma}\rm}

\newtheorem{proposition}[theorem]{{\sc  Proposition}\rm}
\newtheorem{Proposition}[theorem]{{\sc  Proposition}\rm}
\newtheorem{remark}[theorem]{{\sc Remark}\rm}

\begin{document}

\title{The existence theorem for the general relativistic Cauchy problem on the light-cone%
\thanks{Preprint UWThPh-2012-29}}
\author{
Piotr T. Chru\'sciel\thanks{Email  {
Piotr.Chrusciel@univie.ac.at}, URL {
http://homepage.univie.ac.at/piotr.chrusciel/}}
\\
\\{I.H.\'E.S., Bures sur Yvette}\\ and\\ {University of Vienna}
}
\maketitle
\begin{abstract}
We prove existence of solutions of the vacuum Einstein equations with initial data induced by a smooth metric on a light-cone.
\end{abstract}

\tableofcontents

\section{Introduction}

 \newcommand{\fl}{}
\newcommand{\og}{{\overline{g}}}
\newcommand{\zH}{{\mathring{H}}}
\newcommand{\hGamma}{{\hat{\Gamma}}}
\newcommand{\cxi}{\,{{\check{\!\xi}}}{}}
\newcommand{\Oiri}{O_\infty(r^{\infty}){}}
\newcommand{\checkC}{{\check C}{}}

\newcommand{\Rell}{{\overset{(\ell)}R}{}\! }
\newcommand{\psilp}{{\overset{(\ell )}\psi}{}\! }
\newcommand{\dpsilp}{{\overset{(\ell )}{\delta\psi}}{}\!\!}
\newcommand{\psil}{{\overset{(\ell-1)}\psi}{}\!\!}
\newcommand{\chil}{{\overset{(\ell-1)}\chi}{}\!\!}
\newcommand{\dchilp}{{\overset{(\ell )}{\delta\chi}}{}\!\!}
\newcommand{\chilp}{{\overset{(\ell )}\chi}{}\!\!}
\newcommand{\gello}{{\overset{(1)}g}{}\!\!}
\newcommand{\Rello}{{\overset{(1)}R}{}\!\!}
\newcommand{\Rellz}{{\overset{(0)}{R}}{}\!\!}
\newcommand{\gellz}{{\overset{(0)}{g}}{}\!\!}
\newcommand{\xellz}{{\overset{(0)}x}{}\!}
\newcommand{\gell}{{\overset{(\ell)}g}{}\! }
\newcommand{\dxell}{\delta {\overset{(\ell+3)}x}{}\!}
\newcommand{\xell}{{\overset{(\ell)}x}{}\!}
\newcommand{\dgell}{{\overset{(\ell+2)}{\delta g}}{}\!\!\!}
\newcommand{\dgellm}{{\overset{(\ell+1)}{\delta g}}{}\!\!\!}
\newcommand{\Gammal}{{\overset{(\ell)}\Gamma}{}\! }
\newcommand{\Gammalp}{{\overset{(\ell+1)}\Gamma}{}\!\! }
\newcommand{\Rellp}{{\overset{(\ell+1)}R}{}\!\!\!}
\newcommand{\gellp}{{\overset{(\ell+1)}g}{}\!\!\!}
\newcommand{\xellp}{{\overset{(\ell+1)}x}{}\!}
\newcommand{\gellm}{{\overset{(\ell-1)}g}{}\!\!\!}
\newcommand{\gone}{{\overset{(1)}{g}}{}\!\!}

\newcommand{\usigiy}{\overset{(i+2)}\underline{\sigma}_{\!\!\! jk}}
\newcommand{\usigi}{\overset{(i+2)}{\underline{\sigma}}_{\!\!\! AB}}
\newcommand{\divsigi}{{\mathcal D}^j \overset{(i+2)}{\underline{\sigma}}_{\!\!\! j k}}
\newcommand{\fif}{\overset{(i+2)}f}
\newcommand{\gi}{\overset{(i+2)}g}
\newcommand{\fm}{\overset{(m)}f}
\newcommand{\fmpt}{\overset{(m+2)}f}
\newcommand{\gmpt}{\overset{(m+2)}g}
\newcommand{\usigmy}{\underline{\overset{(m)}\sigma}_{\!\!ij}}
\newcommand{\usigm}{\underline{\overset{(m)}\sigma}_{\!\!\! AB}}
\newcommand{\divsigm}{{\mathcal D}^j \underline{\overset{(m)}\sigma}_{\!\!\! ij}}

\newcommand{\myw}{x}
\renewcommand{\regular}{y}

\newcommand{\ccheck}[1]{{\check{#1}}}

\newcommand{\whC}{{\check C}}

A systematic way of constructing general solutions of the vacuum Einstein equations proceeds via solving of Cauchy problems of various flavors. One such classical problem consists of prescribing initial data on a light-cone.
The formal aspects of this Cauchy problem are well understood by now~\cite{CCM2,ChPaetz,RendallCIVP,DamourSchmidt}. However, because of the singularity at the vertex, there arise significant difficulties when attempting to prove an existence theorem for general initial data, and only special cases have been established in the literature so far~\cite{CCM2,CCM3}. It is the purpose of this work to fill this gap and prove an existence theorem for an exhaustive class of initial data, in the sense that every smooth light-cone in every smooth vacuum space-time arises from our construction.

Now, in order to prove existence of a space-time with initial
data on a light-cone $C_{O}$, with vertex $O$, using the wave-map gauge scheme of~\cite{CCM2,RendallCIVP,DamourSchmidt}, one needs to prove that the fields  $  {g}_{\mu\nu}|_{C_O}$ arise from some smooth metric, so that the Cagnac-Dossa
theorem applies~\cite{DossaAHP}.
In this scheme
the fields  $  {g}_{\mu\nu}|_{C_O}$  are constructed by solving a set of wave-gauge constraint equations starting from geometric initial data $(\tilde g, \kappa)$ (for notation, see below and~\cite{CCM2}), which results in a tensor field $  {g}_{\mu\nu}|_{C_O}$ on $C_{O}$ with seemingly intractable behaviour at the vertex.  The problem addressed, and solved, in this work is to show that  $  {g}_{\mu\nu}|_{C_O}$ is indeed the restriction to $C_O$ of some smooth metric, which leads to our first main result:

\begin{theorem}
 \label{T12II.1x}
Consider a symmetric tensor $\tilde g$ induced by a smooth
Lorentzian metric $C$ on its null cone $C_O$ centred at $O$. Then
there exists a smooth metric $g$ defined in a neighborhood of $O$, solution  of the vacuum Einstein
equations to the future of $O$,  such that $C_O$ is the light-cone of $g$ and   $\tilde g$ is the restriction of $g$ to $C_O$.
%
\end{theorem}

Theorem~\ref{T12II.1x} is obtained from Theorem~\ref{T12II.1} of  Section~\ref{S25VIII12.1} by calculating algebraically $\kappa$, in a neighbourhood of the vertex $O$, in terms of $\tilde g$ and its derivatives, using  \eq{19V10.3xy} below. The regularity of the function $\kappa$ needed in Theorem~\ref{T12II.1} is justified in Section~\ref{s8V12.5}.

In Rendall's approach to the characteristic initial value problem~\cite{RendallCIVP} one requires $\kappa=0$. In adapted coordinates on the light-cone, one prescribes a tensor field $\gamma_{AB}(r,x^C)\,dx^A dx^B$ which   determines $\tilde g$ after multiplication by a conformal factor. In this context we prove:

\begin{theorem}
 \label{T12II.1R}
Let
$$\gamma_{AB}(r,x^C)\,dx^A dx^B
$$
be  induced by a smooth Lorentzian metric $C$ on its light-cone centred at $O$ in adapted coordinates, where $r$ is a $C$-affine parameter. Then there exists a
smooth metric  $g$ defined in
a  neighbourhood of $O$, solution of the vacuum Einstein equations in $J^+(O)$, such that
$$
  {g}_{AB}|_{C_O}=\Omega^2\gamma_{AB}
 \;,
$$
for some positive function $\Omega $ which is the restriction to $C_O$ of a smooth function on space-time, where $r$ is a $g$-affine parameter.
%
\end{theorem}

We show in Section~\ref{s10II12.1} how to deduce Theorem~\ref{T12II.1R}
from Theorem~\ref{T12II.1} with $\kappa=0$.

In the scheme of~\cite{ChPaetz}, where the fields $  {g}_{\mu\nu}|_{C_O}$ are given a priori, and the ``wave-map gauge constraint functions" $\Box_g y^\mu|_{C_O}$ are calculated from the data, one needs to prove that a certain vector field $\What{}^\mu$  arises as the restriction to the light-cone of a smooth vector field. We prove that this is the case when the fields $  {g}_{\mu\nu}|_{C_O}$ arise as the restriction to the light-cone of a smooth metric, leading to:

\begin{theorem}
 \label{T29VOOO12/1}
Given any smooth metric $C$ there exists a smooth metric $g$, defined on a neighbourhood of $O$ and solving the vacuum
 Einstein equations to the future of $O$, such that
 $$
  g_{\mu\nu}|_{C_O}= C_{\mu\nu}|_{C_O}
  \; .
 $$
\end{theorem}

The proof of Theorem~\ref{T29VOOO12/1} is the contents of Section~\ref{S28VII12.1}.

\section{Outline of the argument}
 \label{S9V12.10}

Throughout we use the conventions and notations from~\cite{CCM2}.
In particular the coordinates $x$ are linked to the coordinates $y$, which
define $\R ^{n+1}$ as a $C^{\infty }$ manifold, by the relations
\begin{equation}
y^{0}=x^{1}-x^{0},\mbox{ \ \ }y^{i}=r\Theta ^{i}(x^{A})\mbox{ \ with \ }%
\sum_{i=1}^{n}\Theta ^{i}(x^{A})^{2}=1,
 \label{19XI.2}
\end{equation}
the $x^{A}$ are local coordinates on the sphere $S^{n-1}$, or angular polar
coordinates.  We underline components of tensors in coordinates $y$ and
don't underline those in coordinates $x$;  we overline the restrictions to (``traces on") $C_{O}$.  Thus $g_{\mu\nu}$ denotes the components of the metric in the $x$--coordinate system, $\underline{g_{\mu\nu}}$ or $\underline{g}{}_{\mu\nu}$ denotes the components of the metric in the $y$--coordinate system, $\overline{ g}_{\mu\nu}$ denotes the restriction to the light-cone of the components of the metric in the $x$--coordinate system, etc. We use the wave-map gauge with a
Minkowskian target metric; in the notation of \cite{CCM2}, $\widehat g \equiv\eta$.

We assume that we are given a smooth metric $C$, for which we
introduce normal coordinates $\regular^\mu$.  As discussed in~\cite{ChPaetz},
there are many ways in  which $C$ can be used  for the construction of a solution.
One of the schemes analysed here
 assumes that $C_{AB}$ provides the initial data tensor field $\tilde g:=\overline{g}_{AB}dx^A dx^B$
directly,
$$
 \overline{g}{}_{AB}:=\overline{C}_{AB}
  \;,
$$
in which case the parallel-transport coefficient $\kappa$ is determined, at least in a neighborhood of the vertex $O$, from
$\overline{C}_{AB} $ by algebraically solving the Raychaudhuri equation. In a second scheme considered here the metric functions $C_{AB}$ provide a conformal class,
$$
 \overline{g}{}_{AB}:=\Omega^2 \overline{C}_{AB}
 \;,
$$
in which case we solve the Raychaudhuri equation, understood as a second-order ODE for the conformal factor $\Omega$.
We show how the second scheme can be reduced to the first.
After this, the objective is  to construct all the metric functions
$\overline{g}{}_{\mu\nu}$ on the light-cone, with $y$-coordinate
components $\underline{\overline{g}{}_{\mu\nu}}$
which are restrictions to the light-cone of  functions which are smooth in the coordinate system $\regular^\mu$,
by solving the wave-map gauge constraint equations of~\cite{CCM2}. As already pointed out, the
difficulty is to show cone-smoothness%
\footnote{We say that a function $f$ on $C_O$ is \emph{cone-smoth} if there exists a smooth function $\phi$ on space-time such that $f$ is the restriction of $\phi$ to  $C_O$.}
 of the metric functions
$\underline{\overline{g}{}{_{\mu\nu}}}$ near the tip of the
light-cone.

Finally, we consider a scheme where all the metric functions are prescribed directly on $C_O$ using the metric $C$,
$$
 \overline{g}{}_{\mu\nu}:=  \overline{C}_{\mu\nu}
 \;.
$$
In this case the equations $\overline S_{\mu\nu}\ell^\nu=0$, where $S$ is the Einstein tensor and $\ell$ is tangent to the generators of the light-cone, become equations for a wave-gauge vector $\overline H ^\mu$. The
difficulty is then to show cone-smoothness
 of the metric functions
$\underline{\overline H ^\mu}$ near the tip of the
light-cone.

The argument can be outlined as follows: In the affinely
parameterized case,
 where only the conformal class of
$\overline{g}{_{AB}}$ is given, we first analyze the scalar
$|\sigma|^2$, which depends only on the conformal class
of the angular block of the metric. Using the first constraint (the Raychaudhuri equation),
we determine the divergence $\tau$ and the conformal factor
$\Omega$ relating $\overline{g}{_{AB}}$ and the initial data
$\gamma_{AB}\equiv \overline{C}_{AB}$, and analyze their
properties at the vertex. This part of the argument is rather similar to that in~\cite{CCM4} where, however, restrictive hypotheses
have been made on the initial data.

In the case where $\overline{C}{_{AB}}$ gives directly
$\overline{g}{_{AB}}$, instead of the above we determine
algebraically $\kappa$ in a neighborhood of the vertex from $\overline{C}{_{AB}}$ using the first
constraint equation.

The next key step, established in Section~\ref{S3VI0.1}, is
the proof of existence of a smooth space-time metric
$\checkC _{\mu\nu}$, in wave-map gauge,  which solves
all the wave-map gauge constraint equations up to an error term
which, for smooth $C_{\mu\nu}$'s, decays to infinite order%
\footnote{A function $f$ is said to decay to infinite order near $r=0$, we then write $f=\Oiri $, if for all $N\in \N$ we have $|f|\le C_N r^N$ for small $r$ for some constant $C_N$; similarly for all derivatives of $f$.}
at
the origin, with $\overline{\checkC }{_{AB}}=\overline{g}{_{AB}}$,
and with the corresponding function $\check \nu_0$ associated
with $\checkC $ differing from $\nu_0$ by an error term which
again decays to infinite order
at the origin.
For $C_{\mu\nu}$ with finite differentiability, say $C^k$, the
error terms above can be made to  decay to order $O(|\vec
w|^{k-m_1})$, for some $m_1\in \N $ which does neither depend
upon the differentiability index $k$ nor upon the dimension
$n$.
The second constraint equation is then rewritten as an equation
for
$$
 \nu_{A}-\overline{\checkC }{_{0A}}
 \;,
$$
the solution of which is shown to decay to infinite
order at the origin in the smooth case, or to order $O(|\vec
\regular|^{k-m_1-m_2})$, for some $m_2\in \N $ which again does not
depend upon $k$ or $n$ in the $C^k$ case. Similarly the final
constraint is rewritten as an equation for
$$
 \overline{g}{_{00}}-\overline{{\checkC }}{_{00}}
 \;,
$$
the solution of which has similar decay properties at the
origin. The decay properties of the  differences of metric
functions allow one to show that the $\regular^\mu$--coordinate
components $\underline{\overline{ g_{\mu\nu}}}$ of $\overline{
g }$ can be smoothly extended off the light-cone in the
$C^{\infty}$ case, or $C^{k-m_1-m_2-m_3}$-extended in the $C^k$
case, for some $m_3$ independent of $k$ and $n$. This allows us
to use the Cagnac-Dossa theorems~\cite{Cagnac1981,Dossa97} to
solve the wave-gauge reduced Einstein equations, and the
results in~\cite{CCM2} lead to Theorem~\ref{T12II.1} below for
all $k$ large enough.

In the ``unconstrained" scheme of \cite{ChPaetz} where the whole metric is prescribed on the light cone,
$$
 \overline g_{\mu\nu}= \overline C_{\mu\nu}
 \;,
$$
we use similarly a metric $\checkC _{\mu\nu}$ which now is \emph{not} assumed to be in wave-map gauge, but the Ricci tensor of which, when contracted with a null tangent to the light-cone, decays to infinite order on $C_O$ near the vertex along the light-cone.
Comparing a suitably defined gauge vector $\overline H{}^\mu$, as calculated for the desired vacuum metric, with the harmonicity vector $\overline{\check H}{}^\mu$ as calculated for the metric $\checkC _{\mu\nu}$, allows us to show that $\ol H{}^\mu$ extends smoothly.

\section{From a conformal class $\gamma$ to $\tilde g$}
 \label{s10II12.1}

Consider a tensor field  $\gamma$ which is induced on
$C_{O}$ by a smooth metric $C $  in a spacetime neighbourhood
of $C_{O}$, i.e. $\gamma _{AB}=\overline{C}_{AB}$, with
$C_{AB}$ the components with indices $AB$ in the coordinates
$x$ of a metric $C$ whose smoothness is insured by the
smoothness of its components $\underline{C_{\alpha\beta}}$ in
the $\regular$ coordinates. (This property is clearly a
necessary   condition for the desired vacuum metric to satisfy the requirements of our theorem.) Then $\overline{g}{}_{AB}=\Omega^2 \gamma_{AB}$ will be the components in
the coordinates $x$ of indices $AB$ of the trace of a smooth
metric if the conformal factor $\Omega$ is the trace of a smooth
positive function.

Consider a metric $C$ such that $C(O)=\eta$, the Minkowski
metric.
If $C$ is $C^{k}$ with values $\eta $ at $O$ then its
components in the coordinates $\regular$ admit at $O$ an
expansion, where the $c$'s are numbers, and the error terms
$o_{k}(|\regular|^{k})$  (see the beginning of
Appendix~\ref{ss8VIII0.1} for the definition of the symbol
$o_{k}(|\regular| ^{k})$) are $C^{k}$ functions of the
$\regular$'s, of the form
\begin{equation*}
\underline{C_{\alpha\beta}}=\underline{\eta_{\alpha\beta}}%
+\sum_{p=1}^{k}\frac{1}{p!}c_{\alpha\beta,\alpha_{1}\cdots\alpha_{p}}
\regular^{\alpha_{1}}\cdots \regular^{\alpha_{p}}
+o_{k}(|\regular| ^{k})
\end{equation*}
(compare Lemma~\ref{L23I.1}). If $\underline{\partial_{\alpha
}C_{\beta \gamma }}(O)=0$ the expansion starts at $p=2$. This
is satisfied in particular if the $\regular$'s are normal
coordinates for $C$ with origin $O$. In the coordinates $x$ it
holds that
\begin{equation*}
C_{AB}\equiv r^{2}\underline{C_{ij}}\frac{\partial\Theta^{i}}{\partial x^{A}}
\frac{\partial\Theta^{j}}{\partial x^{B}}\;,
\end{equation*}
where
$$
 \mbox{$\regular^{0}=r$,\
$\regular^{i}=r\Theta^{i}$, \  $\displaystyle \sum_{i=1}^n(\Theta^{i})^{2}=1$.}
$$
On $C_O$ this leads to an expansion of the form, with $c$'s
and $d$'s  numbers determined by the $\underline{C_{ij}} $'s:
\begin{eqnarray}
 \nonumber
 \gamma_{AB}&\equiv &
 \overline{C}{_{AB}}
\\
 \nonumber
 &= & r^{2}\Big\{
s_{AB}
+\frac{\partial\Theta^{i}}{\partial x^{A}}
 \frac{\partial\Theta^{j}}{\partial x^{B}}
\sum_{p=1}^{k}\big(
      c_{ij,h_{1}...h_{p}}\regular^{h_{1}}\cdots \regular^{h_{p}}
\\
 \nonumber
    &&
    +r\,d_{ij,h_{1}...h_{p-1}} \regular^{h_{1}}\cdots \regular^{h_{p-1}}
     \big)
    \Big\}
    +o_k(r^{k+2})
\\
 & =: &
 r^2 \left\{s_{AB}
 +\frac{\partial\Theta^{i}}{\partial x^{A}}
 \frac{\partial\Theta^{j}}{\partial x^{B}} (c_{ij}+r\,d_{ij})\right\}
 \; .
  \label{10VII0.1}
\end{eqnarray}
An exhaustive intrinsic description of such tensors $\gamma_{AB}$ in space-time dimension four can be found in \cite{ChJezierskiCIVP}.%
\footnote{We take this opportunity to point out that the ``only if" part of Theorem~1.2 of \cite{ChJezierskiCIVP} is not sufficiently justified. However, the ``if" part is correctly proved, and this is enough to infer Theorem~1.1 of \cite{ChJezierskiCIVP}, which is the main result there, and which is what is relevant for the current work.}

\subsection{The functions $|\protect\sigma|^{2}$ and  $\tau$}
 \label{ssfssquare}

The function $|\sigma|^{2}$ which appears in vacuum as a
source of the Einstein wave-map gauge constraints is defined on $C_{O}$ by
\begin{equation*}
|\sigma|^{2} :=\sigma_{A}{}^{B}\sigma_{B}{}^{A},
\end{equation*}
where $\sigma_{A}{}^{C}$ is the traceless part of
$ \frac{1}{2}\gamma^{BC}\partial_{1}\gamma_{AB}$. We assume that there exists a smooth metric $C$ such that
\begin{equation*}
\gamma_{AB}=\overline{C}{_{AB}},
\end{equation*}
and we start by studying the differentiability properties of possible  extensions of $|\sigma|^2$ off the light-cone.

More precisely, let $\regular^\mu$ denote normal coordinates for $C$ centred at $O$.
Set
\bel{29IV12.1}
 L := \regular^\mu \frac{\partial}{\partial \regular^\mu}\;, \quad
 X_{\mu\nu} :=\frac{1}{2}{\cal L}_L C_{\mu\nu} - C_{\mu\nu}
 \;.
\ee
Let the coordinates $x^\mu$ be defined as in (\ref{19XI.2}):
\begin{equation}
\regular^{0}=x^{1}-x^{0},
\qquad
\regular^{i}=x^1\Theta^{i}(x^{A})
\qquad \mbox{with} \quad
\sum_{i=1}^{n}\left(\Theta^{i}(x^{A})\right)^{2}=1
 \;.
 \label{19XI.3}
\end{equation}
We write interchangeably $x^1$ and $r$.
As already mentioned, we underline the components of the metric associated
with the coordinate system $\regular^\mu$, e.g.
$$
 \underline {C_{\mu\nu}}:= C(\partial_{\regular^\mu},\partial_{\regular^\nu})\;,
 \quad
  C_{\mu\nu}:= C(\partial_{x^\mu},\partial_{x^\nu})
  \;,
$$
etc.
Recall that in normal coordinates it holds that~\cite{Thomas} (see~\cite[Appendix~B]{ChJezierskiCIVP} for a reference which is easier to access)
\begin{equation}
 \label{19XI.1}
 \underline{C_{\mu\nu}}\regular^\mu = \underline{\eta_{\mu\nu}} \regular^\mu
 \;.
\end{equation}

One has the identity
\begin{equation}
\label{19XI.3x}
L \equiv x^0 \partial_{x^0}+x^1\partial_{x^1}
 \;,
\end{equation}
which implies that on the light-cone we have $\overline{L} = x^1
\partial_1$ and
$$
 \overline{ {\cal L}_L C_{\mu\nu}}  = x^1\partial_1\overline{C}{_{\mu\nu}} + \overline{\delta_\mu^0 C_{0\nu}
  +\delta_\nu^0 C_{0\mu}+ \delta_\mu^1 C_{1\nu}
  +\delta_\nu^1 C_{1\mu}}
  \;.
$$
In particular,
\begin{equation*}
 \overline{ {\cal L}_L C_{AB}} =
 x^1\partial_1\overline{C}{_{AB}} =
 x^1\partial_1\gamma_{AB}
 \;.
\end{equation*}
It follows from the definition (\ref{19XI.3}) that  (\ref{19XI.1}) is equivalent to
$$
 \overline C_{01}=1\;,\quad \overline C_{i1}=0\;,
$$
which is further equivalent to
\begin{equation}
 \label{11II.3}
 \overline C^{01}=1\;,\quad \overline C^{00}=\overline C^{0A}=0\;,\qquad
 \overline{C}^{AB}=\gamma^{AB} \;,
\end{equation}
where $C^{AB}$ are the contravariant components with indices
$AB$ in the coordinates $x$ of the metric $C$. (The last equation \eq{11II.3} is, of course, a consequence of the remaining ones.)
The tensor $X$ defined in \eq{29IV12.1} obeys the key property
\begin{equation*}
\overline{X_{\mu 1}}=0\;.
\end{equation*}
%
%
This allows us to rewrite
\begin{equation}
\frac{1}{2}\gamma^{BC} \partial_1 \gamma_{AB} =
\frac{1}{2} \overline{C^{BC}\partial_1 C_{AB}} =
\frac{1}{r}(\delta_A^C + \overline{C^{BC}X_{AB}}) =
\frac{1}{r}(\delta_A^C + \overline{Z^{C}{}_{A}})
\;,
 \label{3III12.1}
\end{equation}
where we have introduced the smooth spacetime tensor
\begin{equation}
Z^\nu{}_\mu := C^{\nu\lambda}X_{\lambda\mu}
\;.
 \label{3III12.2}
\end{equation}
Hence $\sigma_A{}^B$ can be constructed from the restriction to
the cone of the traceless part of $Z^B{}_A$. We can then calculate the norm $|\sigma|^2$ using $Z$, as follows: We have
\beal{3III12.3}
 &
 \overline{{\mathrm{tr}} Z }=
 \overline{C^{\mu\nu} X_{\mu\nu}}=
 \gamma^{AB}\overline{ X_{AB}}
 \;,
 &
\\
&
  \overline{|Z| ^2}:= \overline{\mathrm{tr} Z^2} =
  \overline{C^{\mu\alpha}C^{\nu\beta} X_{\mu\nu} X_{\alpha\beta}}=
  \gamma^{AB}\gamma^{CD}\overline{X_{AC}}\overline{X_{BD}}
 \;,
 &
\eeal{3III12.4}
which implies that the norm $|\sigma|^{2}$
equals $(x^1)^{-2}\equiv r^{-2}$
times the restriction of a smooth function in space-time to the
light-cone:
\begin{eqnarray}
 |\sigma|^{2}
 &\equiv&
 \frac 1 {r^2}\big(\overline{
        |Z|^2 - \frac 1 {n-1} ({\mathrm{tr}}Z )^2}
        \big)
 \;,
 \label{20XI.1}
\end{eqnarray}
as desired. Incidentally, this equals
$\frac{1}{r^2}\overline{|Z^{\mathrm{TF}}|^2}$, where $Z^{\mathrm{TF}}$
is the trace-free part of $Z$.

Note that $\underline{X_{\mu\nu}}= O(r^2)$ along $C_O$,
which shows that for smooth metrics $C$ in normal coordinates the function $|\sigma|^{2}$
is $O(r^2)$ and has an expansion for any $k$,
up to a factor $r^{-2}$:
\begin{equation}
|\sigma |^{2}\equiv \frac{1}{r^2}\left(
\sum_{p=4}^{k}\sigma_{i_{1}\ldots i_{p}}\regular^{i_{1}}\cdots \regular^{i_{p}}
+r\sigma _{i_{1}\ldots i_{p-1}}^{\prime}\regular^{i_{1}}\cdots \regular^{i_{p-1}}
+o_k(r^{k})\right)
 \;.
  \label{9V12.10}
\end{equation}
This can also be written as
\begin{equation*}
|\sigma |^{2}\equiv
\sum_{p=4}^{k}\sigma_{p}r^{p-2} +o_k(r^{k-2}) ,
\end{equation*}
{with}
\begin{equation*}
 \sigma_{p}:=\sigma_{i_{1}\ldots i_{p}}\Theta ^{i_{1}}\cdots \Theta^{i_{p}}
 +\sigma_{i_{1}\ldots i_{p-1}}^{\prime}\Theta^{i_{1}}\cdots \Theta^{i_{p-1}} .
\end{equation*}

When $C$ is used to prescribe $\gamma$, the function $\tau$ is obtained by integration of one of the wave-map gauge characteristic constraint equations; we return to this in Section~\ref{ss11II.3}. On the other hand, if $C$ is used to prescribe $\tilde g =\overline{g}_{AB}dx^A dx^B$ directly as $ \overline{C}_{AB}dx^A dx^B$, the \emph{divergence} $\tau$ of the horizon,
\bel{8V12.1}
 \tau:= \frac 12 \overline{g}{}^{AB} \partial_r\overline{g}{}_{AB}
 \;,
\ee
(often denoted by
$\theta$ in the literature; cf., e.g.,
\cite{galloway-nullsplitting})
is calculated from $\overline{C}_{AB}$. In that last case, it follows from \eq{3III12.1}-\eq{3III12.3} that
\begin{equation}
\tau\equiv \frac{1}{2}\gamma^{AB} \partial_1 \gamma_{AB} =
\frac{1}{r}(n-1 + \gamma^{AB}X_{AB}) =
\frac{1}{r}(n-1 + \overline{\tr Z})
\;.
 \label{3III12.5}
\end{equation}
Hence, in such a context the function $r\tau$
is the restriction to the light-cone of a smooth space-time function, with
\bel{30IV12.3}
 r\tau -(n-1) = O(r^2)
 \;.
\ee

In vacuum, the function $\tau$ has to satisfy (cf., e.g.,~\cite{CCM2}) the Raychudhuri equation $\overline{R_{11}}\equiv \overline{R_{\mu\nu}} \ell^\mu
\ell^\nu =0$, where $\ell^\nu$ is a null tangent to the
generators of $C_O$:
\begin{equation}
\partial_{1}\tau -\kappa \tau
+ \frac{\tau^{2}}{n-1}
+ |\sigma|^{2} 
=0
\;,
 \label{19V10.3-1}
\end{equation}
where $\kappa$ has been defined in \eq{8V12.3}.

\subsection{Boundary conditions on $\kappa$}
 \label{s8V12.5}

As discussed in detail in~\cite{CCM2,ChPaetz}, one of the important objects appearing in the formulation of the characteristic initial value problem is   the following connection coefficient:
\bel{8V12.3}
 \nabla_{\partial_r}\partial_r = \kappa \partial_r
 \;.
\ee
A rather natural \emph{gauge-choice} is to assume that the generators of the light-cones are affinely parameterized, which translates to the condition $\kappa=0$. However, it might be more convenient in some situations not to impose this restriction. The question then arises, what is a  natural class of functions $\kappa$ for the problem at hand.

To motivate our hypotheses suppose, momentarily, that the tensor field
$
 \tilde g =\overline{g}{}_{AB}dx^A dx^B
$
arises from a \emph{smooth vacuum} metric $g$, using a smooth coordinate system in which the light-cone takes the usual form $\{y^0 = |\vec y|\}$, but the coordinates $y^\mu$ are not necessarily normal, and so $\kappa$ is not necessarily zero. In this
case $\tau$ still behaves as $(n-1)/r$ near and away from $r=0$,
hence is nowhere vanishing for $r$ sufficiently small. We can then algebraically solve
for $\kappa$ from \eqref{19V10.3-1}:
\begin{eqnarray}
 \kappa &=& \frac 1 \tau\left(\partial_{1}\tau
+ \frac{\tau^{2}}{n-1}
+ |\sigma|^{2} 
 \right)
 \;.
 \label{19V10.3x}
 \end{eqnarray}
We rewrite this in the following way:
\begin{eqnarray}
r\kappa &=& \frac{1}{(r\tau)} \left( r\partial_1(r\tau)
-(r\tau) +\frac{(r\tau)^2}{n-1} + r^2|\sigma|^2 \right)
  \;.
 \label{5III12.1}
\end{eqnarray}
Now we have seen that, in normal coordinates, $r\tau$ and $r^2 |\sigma|^2$ are restrictions to the light-cone of smooth functions, with $r\tau\to_{r\to 0} n-1$. Since $\tau$ and $\sigma$ are intrinsic objects on $C_O$, it is natural to suppose that these properties will remain true in the new coordinates.
This motivates the condition that $r\kappa$ is the restriction to the light-cone of a
smooth function on space-time; equivalently,
\bel{7V12.1}
 \mbox{$r\kappa$ is cone-smooth.
 }
\ee
We also find that
\bel{3IV12.2}
 \kappa = O(r )
 \;,
\ee
whenever $|\sigma |^2 = O(r^2)$  together with \eq{30IV12.3} hold. 

Yet another hint, that \eq{7V12.1}-\eq{3IV12.2} are adequate assumptions on $\kappa$ in many situations, is provided by the following: Suppose that $\tau$ and $\sigma$ arise from the light-cone of a smooth metric $C$, not necessarily vacuum. In normal coordinates $y^\mu $ for $C$ we then have
\begin{equation}
\partial_{1}\tau
+ \frac{\tau^{2}}{n-1}
+ |\sigma|^{2} + \overline{T}_{11}
=0
\;,
 \label{7V12.5}
\end{equation}
where $r\tau$ and $r^2 \sigma^2$ arise by restriction of smooth functions on space-time, and where
$$
 T_{11} = \underline{T_{00}} + 2 \underline{T_{0i}}\frac{y^i} {|\vec y|}  + \underline{T_{ij}}\frac{y^i} {|\vec y|} \frac{y^j} {|\vec y|}
 \;.
$$
If we use $\tilde g:=\overline C_{AB}dx^A dx^B$ as initial data for a vacuum gravitational field, in view of \eq{19V10.3-1} we will have
$$
 \kappa \tau = \overline{\underline{T_{00}} + 2 \underline{T_{0i}}\frac{y^i} {|\vec y|}  + \underline{T_{ij}}\frac{y^i} {|\vec y|} \frac{y^j} {|\vec y|}}
 \;.
$$
Equivalently
$$
 r \kappa  = \frac {1} {r\tau} \big( \overline{\underline{T_{00}}r^2 + 2 \underline{T_{0i}} {y^i}t + \underline{T_{ij}} {y^i}  {y^j} }\big)
 \;,
$$
and so the resulting function $\kappa$ satisfies \eq{7V12.1}-\eq{3IV12.2}.

From now on, consistently with the above, we will assume that the parallel transport coefficient $\kappa$ satisfies  \eq{7V12.1}-\eq{3IV12.2}.

\subsection{Integration of $\tau$ and of the conformal factor}
 \label{ss11II.3}

As already mentioned in the introduction, in the approach of Rendall~\cite{RendallCIVP} the tensor field $\tilde g = \overline g_{AB}dx^A dx^B$ is taken of the form
$$
 \overline g_{AB} = \Omega^2 \gamma_{AB}
 \;,
$$
where the tensor field $\gamma_{AB}$ is a priori given. Equation~\eq{19V10.3-1} becomes then an equation for the conformal factor $\Omega$.

Suppose that $\gamma_{AB}$ arises from a smooth metric $C$: $\gamma_{AB}=\overline C_{AB}$. In the remainder of this section we will show that there exists a smooth positive function on space-time, say $\chi$, so that $\Omega$ is the restriction to the light-cone of $\chi$. Setting $\checkC _{AB}=\chi^2 C_{AB}$, we then obtain $\overline g_{AB}= \overline{\checkC }_{AB}$, where $\checkC $ is a smooth tensor field on space-time. This reduces the study of Rendall's approach to our treatment in Sections~\ref{ss11II.1x}-\ref{S3V12.1} below. To prove existence of $\chi$ we follow the approach in~\cite{CCM3}, with some simplifications, and making more precise the results there, as follows from the current context.

To carry out the analysis it is convenient to introduce
\begin{equation*}
y:=\frac{n-1}{\tau}
 \;,
\end{equation*}
where $\tau$ is the divergence of $C_O$ given by \eq{8V12.1}.
In terms of $y$, the vacuum Raychaudhuri equation~\eq{19V10.3-1}
reads
\begin{equation}
 \label{3V.1}
y^{\prime}=1+ \kappa y + \frac{1}{n-1}|\sigma|^{2}y^{2}
 \;.
\end{equation}
We assume that  $|\sigma|^2$ is of the form~\eq{9V12.10}; this will be true when the  metric $C$ inducing $\gamma$ is $C^{k+1}$, and thus for any $k$ when $C$ is smooth.
We further assume that $\kappa$ satisfies \eq{7V12.1}-\eq{3IV12.2}, with  $r\kappa$ being the restriction to $C_O$ of a function of $C^k$ differentiability class. Lemma~\ref{L23I.1}, Appendix \ref{ss8VIII0.1}, shows that $\kappa$ has an expansion
\begin{equation}
 \label{9V12.11}
\kappa= \frac 1 r \sum_{p=2}^{k}\kappa_{p-1}r^{p}+o_k(r^{k})
 \;,
\end{equation}
with
\begin{equation}
 \label{9V12.13}
\kappa_{p-1}\equiv
\kappa_{i_{1}\ldots i_{p}}\Theta ^{i_{1}}\cdots \Theta ^{i_{p}}
+
\kappa^{\prime}{}_{i_{1}\ldots i_{p-1}}\Theta ^{i_{1}}\cdots \Theta ^{i_{p-1}}
 \;,
\end{equation}
for some collection of numbers $\kappa_{i_{1}\ldots i_{p}}$ and $\kappa_{i_{1}\ldots
i_{p-1}}^{\prime }$.

Using known arguments (compare~\cite{AndChDiss,ChLengardnwe}
and~\cite[Lemma~8.2]{FriedrichCMP86}), it follows from \eq{3V.1} that there exist functions
$ y_i\in C^\infty(S^{n-1})$ such that
\begin{eqnarray}
 \nonumber
 y
 & = &
  \sum_{i=1}^{k+2}y_i r^i + o_{k}(r^{k+2})
\\
  & = &
  r + \frac{\kappa_1}2 r^2%
  +\sum_{i=3}^{k+2}y_i r^i + o_{k}(r^{k+2})
  \label{9V12.15}
\end{eqnarray}
(with the first non-zero term in the sum being equal to $\frac{\sigma_4}5 r^5$ when $\kappa=0$),
where the $y_{p-1}$'s take the form
\begin{equation}
 \label{9V12.14}
y_{p-1}\equiv
y_{i_{1}\ldots i_{p}}\Theta ^{i_{1}}\cdots \Theta ^{i_{p}}
+
y^{\prime}{}_{i_{1}\ldots i_{p-1}}\Theta ^{i_{1}}\cdots \Theta ^{i_{p-1}}
 \;,
\end{equation}
for some collection of numbers $y_{i_{1}\ldots i_{p}}$ and $y_{i_{1}\ldots
i_{p-1}}^{\prime }$. Lemma~\ref{L23I.1} shows that, for all $k\in \N\cup\{\infty\}$,
the function $y/r$ is the restriction to $C_O$ of a $C^k$ function on space-time equal to one at the origin.

Let $\delta y$ be defined as
\begin{equation*}
 y = r(1+\delta y)\;,
\end{equation*}
thus $\delta y$ is the restriction   to $C_O$ of a $C^k$ function on space-time vanishing at the origin.  Hence
\begin{equation*}
 \tau = \frac{n-1}y = \frac{n-1}{r(1+\delta y)}=
        \frac{n-1}{r}\left( 1 - \frac{\delta y} {1+\delta y} \right)
 \;,
\end{equation*}
which shows that $r\tau$ is the restriction to $C_O$ of a $C^k$ function on space-time equal to $n-1$  at the origin.

Let us write
\begin{equation}
 \label{8V.2}
\overline{g}{}_{AB}=e^{\omega} \gamma _{AB}\; ,
\end{equation}
We then have
\begin{equation}
 \label{5VI.5}
\tau=
\partial_{1}\log \sqrt{\det \gamma}
+\frac{n-1}{2} \partial_{1}\omega\; ,
\end{equation}
with $\omega|_{r=0}=0$. Integrating this equation for $\omega$, Lemma~\ref{L1V12.1} allows us to assert that:

\begin{Proposition}
 \label{P9V12.1}
Let $k\in \N\cup\{\infty\}$. Suppose that the metric $\gamma_{AB}$ arises by restriction to $C_O$ of a $C^{k+1}$ metric in normal coordinates, and that we are given a function $r\kappa$ which is the restriction to $C_O$ of a $C^{k}$  function vanishing at the origin to order two. Then the conformal factor $\Omega^2$, relating $\overline g_{AB}$ and $\gamma_{AB}$, obtained by solving the vacuum Raychaudhuri equation,
\begin{equation}
\partial_{1}\tau -\kappa \tau
+ \frac{\tau^{2}}{n-1}
+ |\sigma|^{2} 
=0
\;,
 \label{19V10.3}
\end{equation}
is the restriction to $C_O$ of a $C^{k}$  function which equals one at the vertex.
\qed
\end{Proposition} 

\section{Integration of $\nu_0$}
 \label{ss11II.1x}

From \cite{CCM2}, in vacuum and in wave-map gauge  the following equation has to hold: %
\begin{equation}
 \partial_1\nu^{0}=-\left(\frac{\tau}{2}+\kappa\right)\nu^{0}+ \frac{1}{2}\overline{g}{}^{AB}rs_{AB}
  \;.
 \label{5VI.1311II}
\end{equation}

We want to show that the function $\nu_0$, solution of \eqref{5VI.1311II}, is the restriction
 to the light-cone of a smooth function on space-time. For this we rewrite \eqref{5VI.1311II} as
\begin{equation}
 r\partial_1\nu^{0}=-\left(\frac{r\tau}{2}+r\kappa\right)\nu^{0}+ \frac{1}{2}\overline{g}{}^{AB}r^2s_{AB}
  \;.
 \label{5VI.1311IIxyz}
\end{equation}
Let the conformal factor $\Omega$ be defined by
\begin{equation}
 \label{8V.2new11II}
\overline{g}{}_{AB}=\Omega^2 \gamma _{AB}\; ,
\end{equation}
with $\Omega=1+O(r^2)$, and let $\varphi$ be defined as
\begin{equation}
 \label{10II.111II}
\varphi :=
\left(\frac{\det \tilde{g}}{\det s_{n-1}}\right)^{1/(2n-2)} =
 \Omega \, \left(\frac{\det\gamma}{\det s_{n-1}}\right)^{1/(2n-2)}
 \;,
\end{equation}
with $\varphi = r + O(r^3)$; recall that
\begin{equation}
 \tau = (n-1) \, \partial_{1}\log\varphi ,
 \quad \mbox{equivalently} \quad
 \partial_1 \varphi = \frac{\tau}{n-1}\varphi \; .
 \label{tauvarphi11II}
\end{equation}
We assume first that $\kappa=0$. Using $\varphi$ we can rewrite \eqref{5VI.1311II} in the form
\begin{equation}
 \partial_1(\nu^{0}\varphi^{(n-1)/2})= \frac{\varphi^{(n-1)/2}}{2}\overline{g}{}^{AB}rs_{AB}
  \;,
 \label{5VI.1411II}
\end{equation}
and hence, since $\nu^{0}\varphi^{(n-1)/2}\to_{r\to0}0$,
\begin{equation}
  \nu^{0}(r,x^A)
  = \frac{\varphi^{-(n-1)/2}(r,x^A)}{2}\int_0^r \left(\varphi^{(n-1)/2}\overline{g}{}^{AB}rs_{AB}\right)(s,x^A)\, ds
  \;.
 \label{11II.1}
\end{equation}
From \eqref{11II.3} one has
\begin{equation}
 \label{8VIII0.1}
 \overline{C^{\mu\nu}\eta_{\mu\nu}} = \gamma^{AB}r^2 s_{AB} +2
 \;,
\end{equation}
Using \eq{8VIII0.1} one is led to
\begin{equation}
  \nu^{0}(r,x^A)
  = \frac{\varphi^{-(n-1)/2}(r,x^A)}{2}\int_0^r
  \left(\varphi^{(n-1)/2}\Omega^{-2}(\overline{C^{\mu\nu}\eta_{\mu\nu}}-2)\right)(s,x^A)\, s^{-1}\, ds
  \;.
 \label{11II.4}
\end{equation}
It is then elementary to show (see Lemma~\ref{L1V12.1}, Appendix~\ref{s5IV0.1})  that $\nu^0$ is the restriction to the
light-cone of a smooth function on space-time.
One also finds that $\nu^0\to 1$ as $r$ approaches zero, and closer
inspection of series expansions~\cite{CCM3} shows cancelations which yield
\begin{equation}
 \label{13II.1}
 \nu^0=1+O(r^4)
 \;.
\end{equation}

When $\kappa\ne 0$ we let
\bel{30IV12.1}
 H(r,x^A) = \int_0 ^r  {\kappa(s,x^A)}   ds
 \;,
\ee
and then \eq{11II.4} gets replaced by
\bean
  \nu^{0}(r,x^A)
  & = &  \frac{\left(e^{-H  }\varphi^{-(n-1)/2}\right)(r,x^A)}{2}
   \times
\\
 &&   \int_0^r
  \left(\varphi^{(n-1)/2}\Omega^{-2}(\overline{C^{\mu\nu}\eta_{\mu\nu}}-2)
  e^{H(s,x^A)}\right)(s,x^A)\, s^{-1}\, ds
  \;,
   \qquad
 \label{11II.4v2}
\eea
with identical conclusion.

Summarising:

\begin{Proposition}
 \label{P9V12.2}
Under the hypotheses of Proposition~\ref{P9V12.1}, the solution $\nu_0$ of \eq{5VI.1311II} is the restriction to $C_O$ of a $C^{k}$ function which equals one at the vertex.
\qed
\end{Proposition}

We show in Appendix~\ref{A29VIII12.1} that for any smooth metric $C$ such that $C_{1A}=C_{11}=0$, and for any cone-smooth function $\nu_0$ there exists another smooth metric $\tC$  satisfying $\ol C_{AB}=\ol \tC_{AB}$, $\ol \tC_{1A}=\ol \tC_{11}=0$ and $\ol \tC_{01}=\nu_0$. This is not used in our indirect proof below, but could be used towards a direct proof of our main results in this paper, if such a proof is found.

\section{Approximate polynomial solutions}
 \label{S3VI0.1}

As the next step in our construction, we  construct a smooth metric which is an approximate solution of the constraint equations.

Throughout this section the $x^\mu$'s are Cartesian coordinates on $\R^{n+1}$
in which the metric coefficients are smooth, and the light-cone
is given by the equation $\eta_{\mu\nu}x^\mu x^\nu =0$, where $\eta_{\mu\nu}$ is a diagonal matrix with entries $(-1,1,\ldots,1)$ on the diagonal. This should not be confused with the coordinates adapted to the light-cone,  denoted by $x^\mu$ in the remaining sections of this paper. One can think of the coordinates $x^\mu$ of this section as the coordinates $y^\mu$ of Section~\ref{S9V12.10}, except that we are not assuming that the $x^\mu$'s here are normal for the metric $C$.

\subsection{The scalar wave equation}
 \label{sS3VI0.1}

Let $\Box_\eta$ denote the Minkowskian wave operator,
$$
 \Box_\eta = \eta^{\mu\nu} \partial_\mu \partial_\nu
 \;.
$$
We start with the following observation:

\begin{lemma}
 \label{L2IV10.1}
Let $k\in {\mathbf N}$. For any homogeneous polynomial $P$ of
degree $k$ there exists a unique homogeneous polynomial $W$ of
degree $k+2$ such that $ \Box_\eta W =P$ and $W|_{C_O}=0$.
\end{lemma}

\noindent\proof Any such $P$ can be uniquely written as
\begin{equation}
 \label{2IV10.1}
  P= C_{\alpha_1\ldots \alpha_k} x^{\alpha_1} \cdots x^{\alpha_k}
  \;,
\end{equation}
where $C_{\alpha_1\ldots \alpha_k}$ is symmetric under
permutations,  $C_{\alpha_1\ldots \alpha_k}= C_{(\alpha_1\ldots
\alpha_k)}$: Indeed, the $C_{\alpha_1\ldots \alpha_k}$'s can be
calculated by differentiating  $k$ times the polynomial $P$,
and  hence are unique.

We seek a solution of the form
$$
 W = A_{(\alpha_1\ldots \alpha_k}\eta_{\alpha_{k+1} \alpha_{k+2})} x^{\alpha_1} \cdots x^{\alpha_{k+2}}
 \;,
$$
where $A_{ \alpha_1\ldots \alpha_k} $ is also symmetric in all
indices. All such polynomials $W$ vanish on the light-cone, as
desired.

We start by noting that the map
\begin{equation}
 \label{2IV10.4} A_{ \alpha_1\ldots \alpha_k} \mapsto W =
 A_{(\alpha_1\ldots \alpha_k}\eta_{\alpha_{k+1} \alpha_{k+2})}
 x^{\alpha_1} \cdots x^{\alpha_{k+2}}
\;,
\end{equation}
which is surjective by definition, is also injective.  Indeed,
this statement is equivalent to the fact that the only solution
of the equation
\begin{equation}
 \label{3IV10.1x}
 A_{(\alpha_1\ldots \alpha_k}\eta_{\alpha_{k+1}
 \alpha_{k+2})}=0
 \;,
\end{equation}
is zero. To see this, let $k+2=2m+\epsilon$, with $\epsilon \in
\{0,1\}$. Contracting \eqref{3IV10.1x} with $\eta^{\alpha_1
\alpha_2} \ldots \eta^{\alpha_{2m-1}\alpha_{2m}}$ we find
$$
0 = \left\{
   \begin{array}{ll}
 A^{\alpha_1}{}_{\alpha_1}{} \ldots^{\alpha_m}{}_{\alpha_m}{}, & \hbox{$k=2m$;} \\
 A^{\alpha_1}{}_{\alpha_1}{} \ldots^{\alpha_m}{}_{\alpha_m \alpha} , & \hbox{$k=2m+1$.}
   \end{array}
 \right.
$$
If $m$ equals zero or one we are done. Otherwise one can
contract now \eqref{3IV10.1x} with $\eta^{\alpha_1 \alpha_2}
\ldots \eta^{\alpha_{2m-3}\alpha_{2m-2}}$, and using the
previous equation obtain
$$
0 = \left\{
   \begin{array}{ll}
 A^{\alpha_1}{}_{\alpha_1}{} \ldots^{\alpha_{m-2}}{}_{\alpha_{m-1} \beta \gamma}{}, & \hbox{$k=2m$;} \\
 A^{\alpha_1}{}_{\alpha_1}{} \ldots^{\alpha_{m-2}}{}_{\alpha_{m-1} \beta \gamma \delta}{} , & \hbox{$k=2m+1$.}
   \end{array}
 \right.
$$
Continuing this way, after a finite number of steps we obtain
the vanishing of $A_{\alpha_1\ldots \alpha_k}$, as desired. 

Consider, now, the linear map which to the tensor $A_{
\alpha_1\ldots \alpha_k} $ assigns the tensor
$C_{\alpha_1\ldots \alpha_k}$, obtained  in the obvious way
from what has been said so far:
$$
 A_{\alpha_1\ldots \alpha_k} \longleftrightarrow
 W\mapsto \Box_\eta W\longleftrightarrow C_{\alpha_1\ldots \alpha_k}
 \;.
$$
This map is injective: indeed, let $\Box_\eta W_2 =P=\Box W_1$,
then $\Box_\eta (W_1-W_2)=0$, with $W_1-W_2=0$ on the
light-cone, hence $W_1-W_2=0$ by uniqueness of solutions of the
characteristic Cauchy problem on the light-cone. Surjectivity
follows now by elementary finite-dimensional algebra.
\hfill $\Box$

\bigskip

For further reference we note that for $k\ge 2$ one finds, in
space-time dimension $n+1$,
%
%
\begin{eqnarray}
 \nonumber
  \Box W &= &   (k+2)\eta^{\mu\nu} \partial_\nu\Big( A_{(\alpha_1\ldots \alpha_k}\eta_{\alpha_{k+1} \mu)}
  x^{\alpha_1} \cdots x^{\alpha_{k+1}}\Big)
\\
 \nonumber
   & = &
   (k+2)(k+1) \eta^{\mu\nu}  A_{(\alpha_1\ldots \alpha_k}\eta_{  \mu\nu)}
  x^{\alpha_1} \cdots x^{\alpha_{k }}
\\
 \nonumber
   & = &
  \frac {(k+2)(k+1)} { (k+2)!}\eta^{\mu\nu}\Big(
  2 k!
 A_{ \alpha_1\ldots \alpha_k}\eta_{\mu\nu}
 +
  4 \times k \times k!
 A_{\mu (\alpha_1\ldots \alpha_{k-1}}\eta_{\alpha_k)\nu}
\\
 \nonumber
   &  &
   \phantom{xxxxx}
 +
      k\times (k-1) \times k!
 A_{\mu \nu (\alpha_1\ldots \alpha_{k-2}}\eta_{ \alpha_{k-1} \alpha_{k })}
 \Big) x^{\alpha_1} \cdots x^{\alpha_k}
\\
 \nonumber
 & = &
   \eta^{\mu\nu}\Big(2
  A_{ \alpha_1\ldots \alpha_k}\eta_{\mu\nu}
 +
  4   k
 A_{\mu (\alpha_1\ldots \alpha_{k-1}}\eta_{\alpha_k)\nu}
\\
 \nonumber
   &  &
   \phantom{xxxxx}
 +
    k  (k-1)
 A_{\mu \nu (\alpha_1\ldots \alpha_{k-2}}\eta_{ \alpha_{k-1} \alpha_{k })}
 \Big) x^{\alpha_1} \cdots x^{\alpha_k}
\\
 \nonumber
 & = &
   \Big( 2(n+2k+1)
  A_{ \alpha_1\ldots \alpha_k}
\\
   &  &
   \phantom{xxxxx}
 +
    k  (k-1)
 A^\mu {}_{\mu (\alpha_1\ldots \alpha_{k-2}}\eta_{ \alpha_{k-1} \alpha_{k })}
 \Big) x^{\alpha_1} \cdots x^{\alpha_k}
  \;.
 \label{10IV10.1}
\end{eqnarray}
So Lemma~\ref{L2IV10.1} is equivalent to the statement that the
equations
\begin{eqnarray}
  2(n+2k+1)
  A_{ \alpha_1\ldots \alpha_k}
 +
    k  (k-1)
 A^\mu {}_{\mu (\alpha_1\ldots \alpha_{k-2}}\eta_{ \alpha_{k-1} \alpha_{k })}
  =
  C_{\alpha_1\ldots \alpha_k}
  \;.
 \label{2III10.2}
\end{eqnarray}
have a unique totally symmetric solution $A_{ \alpha_1\ldots \alpha_k}$ for any totally symmetric  $
C_{\alpha_1\ldots \alpha_k}$.

A similar but simpler calculation shows that the formula \eq{10IV10.1}
remains valid for $k=0$ and $1$, and so for example we obtain
%
%
$$
W= \left\{
   \begin{array}{ll}
      \frac C {2 (n+1)} \eta_{\alpha\beta} x^\alpha x^\beta , & \hbox{$k=0$;} \\
       \frac 1 {2 (n+3)} C_{(\gamma} \eta_{\alpha\beta)} x^\alpha x^\beta x^\gamma, & \hbox{$k=1$.}
   \end{array}
 \right.
$$

\bigskip

As an obvious corollary of Lemma~\ref{L2IV10.1} one finds:

\begin{corollary}
 \label{C3IV10.1}
Let $k\in {\mathbf N}$. For any  polynomial $P$ of degree $k$
there exists a unique  polynomial $W$ of degree $k+2$ such that
$ \Box_\eta W =P$ and $W|_{C_O}=0$.
\hfill $\Box$
\end{corollary}

Let $\Box_g$ be Laplace-Beltrami operator of a metric $g$. As a
warm-up, we prove:

\begin{proposition}
 \label{P3IV10.2}
Let $g$ be a smooth Lorentzian metric and let there be given a coordinate system near $p$ such that $x^\mu(p)=0$. For any smooth function
$\psi$  there exists a unique polynomial $\phi_{k+2}$ of
degree $k+2$ such that
\begin{equation}
 \label{3IV10.5}
 \Box_g\phi_{k+2}-\psi = O(|\myw|^{k+1})
 \;, \qquad \phi_{k+2}|_{C_O}=0
 \;.
\end{equation}
If $\psi=O(|\myw|^{\ell})$, then
$\phi_{k+2}=O(|\myw|^{\ell+2})$. The result remains true for
$k=\infty$, in the sense that there exists a smooth function
$\phi_\infty$ vanishing at the light-cone such that
$\Box_g\phi_{\infty}-\psi$ vanishes to arbitrary order at the
origin, similarly for derivatives of arbitrarily high order of
$\Box_g\phi_{\infty}-\psi$.
\end{proposition} 

\noindent\proof By a linear change of coordinates we can
without loss of generality assume that $g(0)=\eta$.

We will use induction upon
$k$.

For $k=0$, existence is obtained by setting $\phi_2 =
\frac {\psi(0)} {2 (n+1)} \eta_{\alpha\beta} \myregular^\alpha
\myregular^\beta$. To prove uniqueness, consider the difference of two such polynomials solving \eq{3IV10.5}, call it $W$. Introduce a new coordinate system where $x^i$ is replaced by $\epsilon x^i$; one obtains
\bel{11III12.1}
 \partial_ i \big(\sqrt{\det g(\epsilon x)} g^{ij}(\epsilon x) \partial_j W(x)\big) = O(\epsilon |x|)
 \;.
\ee
Passing to the limit $\epsilon\to 0$ we find
$$
 \Box_\eta W = 0
 \;,
$$
and since $W$ vanishes on the light-cone, the vanishing of $W$ follows from, e.g., Corollary~\ref{C3IV10.1}.

Suppose, next, that the result has been established for some
$k$, thus there exists a polynomial solution $\phi_{k+2}$ to
\eqref{3IV10.5}.

Taylor expanding $\psi$, we can write
$$
\psi = \psi_{k } + \delta\psi_{k+1 } + O(|\myw|^{k+2})
 \;,
$$
where
$ \psi_{k } $ is a  polynomial of order $k $, and
$\delta\psi_{k +1} $ is a homogeneous polynomial of order
$k+1$. Similarly Taylor expanding $\Box_g\phi_{k+2}$, we can
write
\begin{equation}
 \label{3IV10.6}
 \Box_g\phi_{k+2}-\psi_{k } = \chi_{k+1 }+ O(|\myw|^{k+2})
 \;,
\end{equation}
where $\chi_{k+1 }$ is a homogeneous polynomial of order $k+1$.

Let $\delta\phi_{k+3}$ be the solution given by
Lemma~\ref{L2IV10.1} of the equation
\begin{eqnarray*}
 \Box_\eta\delta\phi_{k+3}  & = & \delta\psi_{k+1 }-\chi_{k+1 }
 \;.
\end{eqnarray*}
This implies
\begin{eqnarray}
 \label{3IV10.8}
 \Box_g\delta\phi_{k+3}  & = & \delta\psi_{k+1 }-\chi_{k+1 } +O(|\myw|^{k+2})
 \;.
\end{eqnarray}
Set
$$
 \phi_{k+3 } = \phi_{k+2 } + \delta\phi_{k+3 }
 \;.
$$
Adding \eqref{3IV10.6} and  \eqref{3IV10.8} we obtain
\begin{equation}
 \label{3IV10.7}
 \Box_g\phi_{k+3}-\psi_{k } -   \delta\psi_{k +1}=O(|\myw|^{k+2})
 \;,
\end{equation}
which implies \eqref{3IV10.5} with $k$ replaced by $k+1$, providing existence of the solution.

Uniqueness follows by a scaling argument similar to the one leading to \eq{11III12.1}, where
the equation for the difference $W$ of two such polynomials becomes instead
\bel{11III12.2}
 \partial_ i \big(\sqrt{\det g(\epsilon x)} g^{ij}(\epsilon x) \partial_j W(x)\big) = O(\epsilon |x|^{k+2})
 \;.
\ee

When $k=\infty$, the function $\varphi_\infty$ is obtained from the above
sequence of polynomials by Borel summation, Lemma~\ref{LBorel} below. Uniqueness
of $\varphi_\infty$ up to a $O(|x|^\infty)$-function follows from what has been said, using the fact that the difference $W$ of any two such solutions satisfies \eq{11III12.2} with an integer $k$ as large as desired.
\hfill $\Box$

\bigskip

The following observation was implicit in the last proof:
\bigskip

\begin{proposition}
 \label{P3IV10.2x}
Let $g$ be a smooth Lorentzian metric, and let $\phi$  be a
smooth function such that, for some $\ell\in \mathrm N$,
\begin{equation}
 \label{3IV10.5a}
 \Box_g\phi  = O(|\myw|^{\ell})
 \;, \qquad \phi|_{C_O}=0
 \;.
\end{equation}
Then
$$\phi=O(|\myw|^{\ell+2})
 \;.
$$
\end{proposition}

\noindent\proof Let $\phi_{k+2}$ be the first
non-vanishing homogeneous polynomial of degree $k+2$ in the
Taylor expansion of $\phi$, and suppose that $k<\ell$. Then
$\phi_{k+2}$ vanishes on $C_O$, and a Taylor expansion of the
left-hand-side of \eqref{3IV10.5a} shows that $\Box_\eta \phi
_{k+2}= 0$, hence $\phi_{k+2}=0$ by Corollary~\ref{C3IV10.1}, a
contradiction.
\hfill \qed

\subsection{The Ricci tensor}
 \label{sS3VI0.2}

We continue with a perturbation lemma, namely: We wish to
deform a given smooth metric $g$ to a new smooth metric $\widehat
g$,  with the property that some components of the Ricci tensor
of $\widehat g$ tend to zero with decay rate $\ell$  along the
light-cone $C_O$ near its tip, with $\ell$ as large as desired,
and such that the new metric coincides with the old one on
$C_O$.

The metric $g$ in the current section should be thought of as the metric $C$ in the remaining parts of the paper. Similarly to Section~\ref{sS3VI0.1}, the symbol  $x^\mu$ is  \emph{not} used to denote the coordinates adapted to the light-cone, as is the case in the main body of the  paper: these are regular space-time coordinates near the vertex in which the light-cone is given by the Minkowskian equation $\eta_{\mu\nu}x^\mu x^\nu=0$.

\begin{lemma}
 \label{3IV10.1} Let $g$ be a smooth Lorentzian metric with the light-cone $C_O$ of $O$ described by the equation $C_O=\{x^\alpha:\ \eta_{\mu\nu}x^\mu x^\nu=0\}$,
where, as elsewhere, $\eta_{\alpha\beta}$ denotes the Minkowski
metric.
We  assume moreover that
%
\begin{equation}
 g_{\alpha\beta} -\eta _{\alpha\beta}= O(|\myw|^2)\;, \quad
 \partial_ \sigma  g_{\alpha\beta} = O(|\myw| )
 \;.
\label{3IV10.2}
\end{equation}
For any $\ell \in\N \cup \{\infty\}$ there exists a
smooth metric $\widehat g$  defined for $|\myw|$ small enough,
which coincides with $g$ on $C_O$,
\begin{equation}
 \overline{g}{_{\mu \nu}}=\overline{\widehat g_{\mu \nu}}
 \;,
 \label{4IV10.4}
\end{equation}
and such that
\begin{equation}
   \widehat R_{\mu\nu}  = O(|x|^{\min{(\ell,2)}})
 \;,
\quad
   \widehat R_{\mu\nu}\myregular^\nu = O(|\myw|^{\ell+1})
    + \overset{(\ell)}P_\mu \eta_{\alpha\beta}\myregular^\alpha \myregular^\beta
 \;,
 \label{4IV10.4b}
\end{equation}
for small $|\myw|$, for some smooth functions
$\overset{(\ell)}P_\nu$, where $\widehat R_{\mu\nu}$ denotes the
Ricci tensor of the metric $\widehat g$, and $\widehat R$ its Ricci
scalar.
\end{lemma}

\noindent\proof The Ricci tensor of $g$ can be written
as
\begin{equation}
R_{\alpha\beta}=-{\frac{1}{2}}\Box_g g_{\alpha\beta}+{\frac{1}{2}}(g_{\alpha\lambda
}\partial_{\beta}\Gamma^{\lambda}+g_{\beta\lambda}\partial_{\alpha}\Gamma^{%
\lambda})+q_{\alpha\beta} (g,\partial g).
\label{RiccihIdentityx}
\end{equation}
Here it is usual to take $\Box_g $ to be the Laplace operator acting on functions,
\bel{13III12.1}
 \Box_g f =
 |\det g|^{-1/2} \partial_\mu(|\det g|^{1/2} g^{\mu\nu}
 \partial_\nu f).
\ee
Further, $q $ is a quadratic form in the first derivatives
$\partial g$ of $g$ with coefficients polynomial in $g$ and its
contravariant associate, and the $\Gamma^{\lambda}$'s  are
defined as
\begin{equation}
 \Gamma^{\alpha}:=g^{\lambda\mu}\Gamma_{\lambda\mu}^{\alpha}
 \;.
\label{Harmonicityx}
\end{equation}
However, instead of \eq{13III12.1} one can  take
$g^{\mu\nu}\partial_\mu\partial_\nu$, with a different $q$ in \eq{RiccihIdentityx}; this implies that it suffices to do the estimates below for $q$ and for $g^{\mu\nu}\partial_\mu\partial_\nu$.

We assume first that $\ell<\infty$. The proof will be done by induction upon $\ell$.

To clarify notation, $\gell$ will denote a metric satisfying
\eqref{4IV10.4b}. We set $\overset{(0)}{g}=g$, consistently with this requirement.
In particular, setting $\overset{(0)}{P}{}_{\!\!\!\mu}=0$, the result is true for $\ell=0$. For $\ell\ge 1$ the metric $\gell$ will be of the form
\begin{equation}
 \label{1VI10.1}
\gell_{\alpha\beta} = \gellm_{\alpha\beta}+\dgellm_{\alpha\beta}
 \;,
\end{equation}
where the correction term $\dgellm_{\alpha\beta}$ will be
$O(|\myw|^{\ell+1})$ near $\myw =0$. Thus, the index $\ell$
over $g$ denotes the induction step, while the index $\ell$
over $\delta g$ denotes the decay rate for small $\myw$. We let
$\Rell_{\alpha\beta}$ denote the Ricci tensor of  $\gell$.

The first step is to achieve the result with $\ell=1$.
In this case the first equality in \eq{4IV10.4b} is the important one, since the second automatically holds with $\overset{(1)}P_\nu=0$. It follows from the calculations that we are about to do that the result is achieved by
setting
\begin{equation}
 \label{5VII0.1}
 \widehat g_{\mu\nu} = \gone := g_{\mu\nu} +
\eta_{\alpha\beta} \myregular^\alpha \myregular^\beta A_{\mu\nu}
 \;,
\end{equation}
where $A_{\mu\nu}$ is given by \eqref{10VI10.7}. The formula \eq{5VII0.1}
defines a Lorentzian metric for $|\myw|$ small enough, and
maintains \eqref{3IV10.2}.

Similarly, for the result  with $\ell=2$ only the first equality in \eq{4IV10.4b} needs to be established, the second one with $\overset{(2)}P_\nu=0$ automatically follows.

In all subsequent steps one wishes to establish the second equality in \eq{4IV10.4b}, making sure that the first one remains true at each induction steps.

So, assuming the result is true for some $\ell\ge 0$,   we  write
\begin{equation}
 \label{1VI10.2}
 \gellp_{\alpha\beta}=\gell_{\alpha\beta}+\dgell_{\alpha\beta}
 \;,
\end{equation}
where $\dgell$ takes the form
\begin{equation}
 \label{9IV10.1}
 \dgell_{\alpha\beta} := A_{\alpha\beta (\gamma_1 \ldots \gamma_{\ell}} \eta_{\gamma_{\ell+1} \gamma_{\ell+2})}
 x^{\gamma_1} \cdots x^{\gamma_{\ell+2}}
 \;,
\end{equation}
and hence vanishes on $C_O$. We consider one by one the terms that occur in
\eqref{RiccihIdentityx} with $g$ there replaced by $\gellp$. We
assume that \eqref{4IV10.4b} holds with $\widehat R_{\alpha\beta}$
replaced by $\Rell_{\alpha\beta}$, and we want to choose
$\dgell_{\alpha\beta}$ to achieve the corresponding properties of the Ricci
tensor of $\gellp$.

The quadratic terms are simplest to analyze:
%
\begin{equation}
q_{\alpha\beta} (\gellp,\partial \gellp)=q_{\alpha\beta} (\gell,\partial \gell) + O(|\myw|^{\ell+2})
 \;.
\label{3IV10.16}
\end{equation}
Indeed, $q$ is a sum of terms of the form
$$
 p(\gellp) \partial \gellp \partial \gellp
 \;,
$$
for a rational function $p$ of $\gellp$, which thus read
(keeping in mind that $\partial \gell =O(|\myw|)$ for all $\ell$)%
%
\begin{eqnarray*}
\lefteqn{
 p(\gell+\dgell\,\,) \partial( \gell+\dgell\,\,) \partial( \gell+ \dgell\,\,)
   = }
 &&
\\
 &&
  \phantom{=\ }
\big(\underbrace{ p(\gell+\dgell\,\,)-p(\gell)}_{ O(|\myw|^{ \ell+2})}\big)
 \underbrace{\partial( \gell+\dgell\,\,)}_{ O(|\myw| )}
 \underbrace{\partial( \gell+\dgell\,\,)}_{ O(|\myw| )}+
 p(\gell) \partial \gell \partial \gell
\\
  &&
  \phantom{=\ }
  +
 2 \underbrace{p(\gell)}_{ O(1 )} \underbrace{ \partial \gell}_{ O(|\myw| )}  \underbrace{\partial \dgell}_{ O(|\myw|^{\ell+1} )} +
 O(|\myw|^{2\ell+2})
\\ &&
  =
 p(\gell) \partial \gell \partial \gell +  O(|\myw|^{\ell+2})
 \;.
\end{eqnarray*}

Now,
\begin{equation}
-{\frac{1}{2}}\Box_{\gellp}{}\,\,{}\,\, \gell_{\alpha\beta}
 = -{\frac{1}{2}}\Box_{\gell}{}\, \gell_{\alpha\beta} + O(|\myw|^{\ell+2})
 \;,
\label{3IV10.13}
\end{equation}
\begin{equation}
-{\frac{1}{2}}\Box_{\gellp}{}\,\,{}\,\, \gellp_{\alpha\beta}
 =  -{\frac{1}{2}}\Box_{\gell}{}\, \gell_{\alpha\beta}    -{\frac{1}{2}}\Box_{\gell}{}\, \dgell_{\alpha\beta}+ O(|\myw|^{\ell+2})
 \;,
\label{3IV10.14}
\end{equation}
where by \eqref{10IV10.1} we also have
\begin{eqnarray}
 \nonumber
 -{\frac{1}{2}}\Box_{\gell}{}\, \dgell_{\alpha\beta}
 &  = &
 -{\frac{1}{2}}\Box_\eta \dgell_{\alpha\beta}   + O(|\myw|^{\ell+2})
\\
 \nonumber
 & = &
   -\frac 12 \Big( 2(n+2\ell+1)
  A_{\alpha\beta \alpha_1\ldots \alpha_\ell} +
    \ell  (\ell-1) \times
\\
   &  &
 A_{\alpha\beta}{}^\mu {}_{\mu (\alpha_1\ldots \alpha_{\ell-2}}\eta_{ \alpha_{\ell-1} \alpha_{\ell })}
 \Big) x^{\alpha_1} \cdots x^{\alpha_\ell} +   O(|\myw|^{\ell+2})
 \;.
 \qquad
  \label{1VI10.4}
\end{eqnarray}
Next,
\begin{eqnarray}
  \nonumber
 \gellp_{\alpha\lambda}\partial_{\beta}\Gammalp^{\lambda}
 &\equiv&
 \gellp_{\alpha\lambda}\partial_{\beta}\Big(\gellp^{\mu\nu}\Gammalp_{\mu\nu}^{\lambda}\Big)
 =
 \gell_{\alpha\lambda}\partial_{\beta}\Big(\gellp^{\mu\nu}\Gammalp_{\mu\nu}^{\lambda}\Big)
 + O(|\myw|^{\ell+2})
\\
\nonumber
 &=&
 \gell_{\alpha\lambda}\partial_{\beta}\Big(\gell^{\mu\nu}\Gammalp_{\mu\nu}^{\lambda}\Big)
 + O(|\myw|^{\ell+2})
\\
 & = &
 \gell_{\alpha\lambda}\partial_{\beta}\Gammal^{\lambda}
 +
 \underbrace{\eta_{\alpha\lambda}\eta^{\mu\nu} \partial_{\beta}\Big(\Gammalp_{\mu\nu}^{\lambda} -\Gammal_{\mu\nu}^{\lambda}\Big)}
 + O(|\myw|^{\ell+2})
 \;,
 \quad
 \phantom{xxxx}
\label{3IV10.15}
\end{eqnarray}
where, to estimate the error term in the last line, we have used
$$
\Gammalp_{\mu\nu}^{\lambda} -\Gammal_{\mu\nu}^{\lambda} =
   O(|\myw|^{\ell+1})
   \;,
    \quad
\partial_\beta\big(\Gammalp_{\mu\nu}^{\lambda} -\Gammal_{\mu\nu}^{\lambda} \big)=
   O(|\myw|^{\ell})
   \;.
$$
The underbraced expression in \eq{3IV10.15} can be analyzed as follows:
\begin{eqnarray*}
 \eta^{\mu\nu} \Big(
 \Gammalp_{\mu\nu}^{\lambda}-\Gammal_{\mu\nu}^{\lambda}\Big)
 & = & \frac 12  \eta^{\mu\nu} \Big( \gellp^{\lambda \sigma} (2\partial _\nu\gellp_{\mu \sigma} - \partial _\sigma\gellp_{\mu \nu})
\\
 &&
 \phantom{xxxxx}
 -\gell^{\lambda \sigma} (2\partial _\nu\gell_{\mu \sigma} - \partial _\sigma\gell_{\mu \nu})
 \Big)
\\
 & = &  \eta^{\mu\nu} \eta^{\lambda \sigma} (\partial _\nu\dgell_{\mu \sigma} - \frac 12 \partial _\sigma\dgell_{\mu \nu})
 + O(|\myw|^{\ell+3})
 \;.
\end{eqnarray*}
The underbraced term in \eqref{3IV10.15} reads thus
$$
 \eta^{\mu\nu}   (\partial_\beta \partial _\nu\dgell_{\mu
\alpha} - \frac 12 \partial_\beta \partial _\alpha\dgell_{\mu \nu})
 + O(|\myw|^{\ell+2})
 \;.
$$
It follows that the sum
$\gellp_{\alpha\lambda}\partial_{\beta}\Gammalp^{\lambda}
 +\gellp_{\beta\lambda}\partial_{\alpha}\Gammalp^{\lambda}$
gives a contribution to the Ricci tensor $\Rellp_{\alpha\beta}$
of $\gellp$ equal to
\begin{eqnarray}
 \label{10IV10.3}
 \frac 12 \eta^{\mu\nu}   (\partial_\beta \partial _\nu\dgell_{\mu
\alpha} +\partial_\alpha \partial _\nu\dgell_{\mu
\beta} -  \partial_\beta \partial _\alpha\dgell_{\mu \nu})
 + O(|\myw|^{\ell+2})
 \;.
\end{eqnarray}
All this leads to the  formula
\begin{eqnarray}
 \nonumber
 \Rellp_{\alpha\beta} & = &
 -{\frac{1}{2}}\Box_\eta \dgell_{\alpha\beta}
  + \frac 12 \eta^{\mu\nu}   (\partial_\beta \partial _\nu\dgell_{\mu
 \alpha} +\partial_\alpha \partial _\nu\dgell_{\mu
 \beta} -  \partial_\beta \partial _\alpha\dgell_{\mu \nu})
\\
&&+ \Rell_{\alpha\beta}  + O(|\myw|^{\ell+2})
 \;.
 \label{1VI10.5}
\end{eqnarray}
For further reference, we note that inserting \eqref{9IV10.1}
with $\ell=0$ into \eqref{10IV10.3} one obtains at $O$
%
\begin{eqnarray}
   2
A_{\alpha \beta  }
 -\eta^{\mu\nu} A_{\mu\nu }
\eta_{\alpha \beta}
 \;.
 \label{10IV10.2}
\end{eqnarray}
Next, the polynomial part of \eqref{10IV10.3} with $ \ell=1$
reads
\begin{eqnarray}
3 \eta^{\mu\nu}   \Big((
A_{\alpha\mu (\beta } +
A_{\beta\mu (\alpha})
\eta_{\nu \gamma)}
 -A_{\mu\nu(\alpha }
\eta_{\beta\gamma)}
 \Big) x^{\gamma}
 \;.
 \label{10IV10.4}
\end{eqnarray}
For $\ell\ge 2$ the corresponding calculations require more
work: We have
\begin{eqnarray*}
 \partial_\beta \partial _\nu\dgell_{\mu
 \alpha}
  & = &
 \bigg(
  \ell (\ell-1) A_{\mu \alpha \beta \nu (\gamma_1 \ldots \gamma_{\ell-2}}
 \eta_{\gamma_{\ell-1}\gamma_\ell)}
 + 2 \ell  A_{\mu \alpha \beta  (\gamma_1 \ldots \gamma_{\ell-1}}
 \eta_{ \gamma_\ell)\nu}
\\
 &&
 + 2 \ell  A_{\mu \alpha \nu  (\gamma_1 \ldots \gamma_{\ell-1}}
 \eta_{ \gamma_\ell)\beta}
 + 2    A_{\mu \alpha \gamma_1 \ldots \gamma_{\ell-1} \gamma_\ell}
 \eta_{\beta\nu  }
 \bigg)
 x^{\gamma_1}\cdots x^{\gamma_\ell}\;,
\\
 \partial_\beta \partial _\alpha\dgell_{\mu
 \nu}
  & = &
 \bigg(
  \ell (\ell-1) A_{\mu\nu \beta  \alpha (\gamma_1 \ldots \gamma_{\ell-2}}
 \eta_{\gamma_{\ell-1}\gamma_\ell)}
 + 2 \ell  A_{\mu \nu \beta  (\gamma_1 \ldots \gamma_{\ell-1}}
 \eta_{ \gamma_\ell)\alpha}
\\
 &&
 + 2 \ell  A_{\mu \nu  \alpha (\gamma_1 \ldots \gamma_{\ell-1}}
 \eta_{ \gamma_\ell)\beta}
 + 2    A_{\mu \nu \gamma_1 \ldots \gamma_{\ell-1} \gamma_\ell}
 \eta_{\beta \alpha }
 \bigg)
 x^{\gamma_1}\cdots x^{\gamma_\ell}
  \;,
\end{eqnarray*}
which results in a polynomial part of \eqref{10IV10.3} equal to 
\begin{eqnarray} \lefteqn{
 \bigg(
  \frac 12 \ell (\ell-1) A^\mu{}_{ \alpha \beta \mu (\gamma_1 \ldots \gamma_{\ell-2}}
 \eta_{\gamma_{\ell-1}\gamma_\ell)}
 + \frac 12 \ell (\ell-1) A^\mu{}_{  \beta\alpha \mu (\gamma_1 \ldots \gamma_{\ell-2}}
 \eta_{\gamma_{\ell-1}\gamma_\ell)}
 +  \ell  A_{ \alpha (\gamma_1 \ldots \gamma_{\ell }) \beta }
 }
 &&
 \nonumber
\\
 &&
 +   \ell  A_{ \beta (\gamma_1 \ldots \gamma_{\ell }) \alpha }
 +   \ell  A^\mu{}_{ \alpha \mu  (\gamma_1 \ldots \gamma_{\ell-1}}
 \eta_{ \gamma_\ell)\beta}
 +   \ell  A^\mu{}_{ \beta\mu  (\gamma_1 \ldots \gamma_{\ell-1}}
 \eta_{ \gamma_\ell) \alpha}
 + 2    A_{\beta \alpha \gamma_1 \ldots \gamma_\ell}
 \nonumber
\\
&& - \frac 12  \ell (\ell-1) A^\mu {}_{\mu  \beta  \alpha (\gamma_1 \ldots \gamma_{\ell-2}}
 \eta_{\gamma_{\ell-1}\gamma_\ell)}
 - \ell  A^\mu {}_{  \mu \beta  (\gamma_1 \ldots \gamma_{\ell-1}}
 \eta_{ \gamma_\ell)\alpha}
 \nonumber
\\
 &&
 - \ell  A^\mu {}_{\mu  \alpha (\gamma_1 \ldots \gamma_{\ell-1}}
 \eta_{ \gamma_\ell)\beta}
 -    A^\mu{}_{ \mu \gamma_1 \ldots  \gamma_\ell}
 \eta_{\beta \alpha }
 \bigg)
 x^{\gamma_1}\cdots x^{\gamma_\ell}
  \;.
 \label{2VI10.1}
\end{eqnarray}

Recall that we  wish to choose $\dgell$ so that the Ricci
tensor of $\gellp$ satisfies \eqref{4IV10.4b} with $\ell$
replaced by $\ell+1$ there. In view of
\eqref{1VI10.4} with $\ell=0$
and \eqref{10IV10.2}, to establish \eq{4IV10.4b} with $\ell=1$ we need to show existence
of solutions to the set of equations
\begin{eqnarray}
-(n-1)A_{\alpha \beta  }
 -\eta^{\mu\nu} A_{\mu\nu }
\eta_{\alpha \beta} = -R_{\alpha \beta}(O)
 \;,
 \label{10IV10.5}
\end{eqnarray}
with symmetric tensors $A _{\alpha \beta} $ and $  R_{\alpha
\beta}(O)$. The solution is
\begin{equation}
 \label{10VI10.7}
  A_{\alpha
\beta} =  \frac 1 {n-1} \left( R_{\alpha
\beta}(O) - \frac 1 {2n} \eta^{\mu \nu}  R_{\mu\nu}(O) \eta_{\alpha \beta}
 \right)
 \;.
\end{equation}

Having thus established the result with $\ell=1$, we expand the Ricci tensor  $\Rellz$ of $\gellz$
in Taylor series to order one,
$$
\Rellz_{\alpha\beta} = C_{\alpha\beta \gamma} \myregular^\gamma + O(|\myw|^2)
 \;  .
$$
In view of the equations derived so far, we will obtain
\begin{equation}
 \label{10VI10.8}
\Rell_{\alpha\beta} =   O(|\myw|^2)
\end{equation}
with $\ell=1$ if we can solve the set of equations
\begin{equation}
 \label{10IV10.11}
 - (n+ 3)
  A_{\alpha\beta \gamma} +
3 \eta^{\mu\nu}   \Big((
A_{\alpha\mu (\beta } +
A_{\beta\mu (\alpha})
\eta_{\nu \gamma)}
 -A_{\mu\nu(\alpha }
\eta_{\beta\gamma)}
 \Big)  = -C_{\alpha\beta \gamma }
 \;,
\end{equation}
keeping in mind that $A$ and $C$ are symmetric in the first two
indices. Moreover, because of the contracted Bianchi identity,
$C$ satisfies
\begin{equation}
 \label{10IV10.6}
 C^\alpha{}_{\alpha \beta} = 2
 C^\alpha{}_{\beta \alpha}
 \;.
\end{equation}

Now,  either directly from \eqref{2VI10.1}, or by expanding,
\eqref{10IV10.11} can be rewritten as
\begin{equation}
 - (n+ 1)
  A_{\alpha\beta \gamma} +A_{\alpha \gamma\beta} +A_{\beta \gamma\alpha} +
 A_{\alpha\mu} {}^\mu  \eta_{\beta\gamma} +
 A_{\beta\mu} {}^\mu  \eta_{\alpha\gamma}
 -3 A^{\mu}{}_{\mu(\alpha }
\eta_{\beta\gamma)} = -C_{\alpha\beta \gamma }
 \;.
 \label{10IV10.12}
\end{equation}

As a consistency check with the contracted Bianchi identity,
we take a trace in $\alpha$ and $\beta$ of \eqref{10IV10.12} to
obtain
%
$$
 -2(n+2) A^\alpha{}_{\alpha \gamma} +  4A_{\gamma\alpha}{}^\alpha = -C^\alpha{}_{\alpha \gamma}
 \;,
$$
while a trace in $\alpha$ and $\gamma$ yields
$$
 -(n+2) A^\alpha{}_{ \alpha \beta} +  2A_{\gamma\alpha}{}^\alpha  = -C^\alpha{}_{\beta\alpha }
 \;,
$$
as required by \eqref{10IV10.6}.

To invert equation \eqref{10IV10.12} we express
$A_{\alpha\beta\gamma}$ as a linear combination of all possible
linear terms which we can form from $C_{\alpha\beta\gamma}$
with the correct symmetry,
with unknown coefficients which need to be determined. Replacing that expression in
\eqref{10IV10.12} gives a linear system for the coefficients,
which we can solve. The result is
\begin{eqnarray}
A_{\alpha\beta\gamma} &=&
\frac{1}{(n+2)(n-1)} \left(
    n C_{\alpha\beta\gamma}
    + C_{\alpha\gamma\beta}
    + C_{\beta\gamma\alpha}
    - C^{\mu}{}_{\gamma\mu} \eta_{\alpha\beta}
 \right)
\nonumber \\ &&
+ c \left(
      C^{\mu}{}_{\beta\mu} \eta_{\alpha\gamma}
    + C^{\mu}{}_{\alpha\mu} \eta_{\beta\gamma}
 \right)
\;,
\end{eqnarray}
where $c$ is an arbitrary constant. Choosing, e.g., $c=0$, establishes our claim with $\ell=1$. 

A similar, but rather more involved, analysis applies for $\ell
\ge 2$; note that \eqref{10VI10.8} remains true under the
current changes of the metric for all $\ell \ge 2$:

We Taylor-expand $\Rell_{\alpha\beta}$ to order $\ell$. Note that so far all error terms were of the order $O(|\myw|^{\ell+2})$, but this Taylor expansion leaves behind an error term $O(|\myw|^{\ell+1})$.   Denote
by
$$   C_{\alpha\beta
\gamma_1 \ldots \gamma_{\ell }}\myregular^{\gamma_1}\cdots  \myregular^{\gamma_{\ell }}
$$
the homogeneous polynomial of order $\ell$ in that Taylor
expansion. In view of \eqref{1VI10.4} and \eqref{2VI10.1}, the
homogeneous polynomial of order $\ell$ in the Taylor expansion
of $\Rellp_{\alpha\beta}$ is 
\begin{eqnarray}
 \nonumber
 &&
 \bigg(-\frac 12 \Big( 2(n+2\ell+1)
  A_{\alpha\beta \gamma_1\ldots \gamma_\ell} +
    \ell  (\ell-1)
 A_{\alpha\beta}{}^\mu {}_{\mu (\gamma_1\ldots \gamma_{\ell-2}}\eta_{ \gamma_{\ell-1} \gamma_{\ell })}
 \Big)
\\
 &&
 +
{
  \frac 12 \ell (\ell-1) A^\mu{}_{ \alpha \beta \mu (\gamma_1 \ldots \gamma_{\ell-2}}
 \eta_{\gamma_{\ell-1}\gamma_\ell)}
 + \frac 12 \ell (\ell-1) A^\mu{}_{  \beta\alpha \mu (\gamma_1 \ldots \gamma_{\ell-2}}
 \eta_{\gamma_{\ell-1}\gamma_\ell)}
 +  \ell  A_{ \alpha (\gamma_1 \ldots \gamma_{\ell }) \beta }
 }
 \nonumber
\\
 &&
 +   \ell  A_{ \beta (\gamma_1 \ldots \gamma_{\ell }) \alpha }
 +   \ell  A^\mu{}_{ \alpha \mu  (\gamma_1 \ldots \gamma_{\ell-1}}
 \eta_{ \gamma_\ell)\beta}
 +   \ell  A^\mu{}_{ \beta\mu  (\gamma_1 \ldots \gamma_{\ell-1}}
 \eta_{ \gamma_\ell) \alpha}
 + 2    A_{\beta \alpha \gamma_1 \ldots \gamma_{\ell-1} \gamma_\ell}
 \nonumber
\\
&& - \frac 12  \ell (\ell-1) A^\mu {}_{\mu  \beta  \alpha (\gamma_1 \ldots \gamma_{\ell-2}}
 \eta_{\gamma_{\ell-1}\gamma_\ell)}
 - \ell  A^\mu {}_{  \mu \beta  (\gamma_1 \ldots \gamma_{\ell-1}}
 \eta_{ \gamma_\ell)\alpha}
 \nonumber
\\
 &&
 - \ell  A^\mu {}_{\mu  \alpha (\gamma_1 \ldots \gamma_{\ell-1}}
 \eta_{ \gamma_\ell)\beta}
 -    A^\mu{}_{ \mu \gamma_1 \ldots  \gamma_\ell}
 \eta_{\beta \alpha }
 +  C_{\alpha\beta
\gamma_1 \ldots \gamma_{\ell }}
 \bigg)
 x^{\gamma_1}\cdots x^{\gamma_\ell}\;.
  \label{2VI10.2}
\end{eqnarray}
Multiplying by $x^\beta$, and disregarding momentarily all
terms involving the Minkowski metric $\eta $ we obtain
\begin{eqnarray}
 &&
 \big(- (n+2\ell-1)
  A_{\alpha\beta \gamma_1\ldots \gamma_\ell}
 +   \ell  A_{ \beta (\gamma_1 \ldots \gamma_{\ell }) \alpha }
 +   \ell  A_{  \alpha  (\gamma_1 \ldots \gamma_{\ell })\beta}
 \nonumber
\\
&&
 \phantom{xxxxxxxxxxxxxxxxxxxxxxxxx}
 +  C_{\alpha\beta
\gamma_1 \ldots \gamma_{\ell }}
 \big)x^{\beta}
 x^{\gamma_1}\cdots x^{\gamma_\ell}\;.
  \label{2VI10.3}
\end{eqnarray}
Set
\begin{eqnarray}
E_{\alpha\beta \gamma_1 \ldots \gamma_{\ell }} & := &
C_{\alpha(\beta \gamma_1 \ldots \gamma_{\ell })}
 \;,
\end{eqnarray}
thus $ E_{\alpha \beta  \gamma_1 \ldots \gamma_{\ell }}$ is
totally symmetric in the last $\ell+1$ indices. Let us write
\begin{equation}
 \label{2VI0.4}
 A_{\alpha\beta \gamma_1 \ldots \gamma_{\ell }}
 = \underbrace{
    a C_{\alpha\beta \gamma_1 \ldots \gamma_{\ell }}
    + b (C_{\alpha  (\gamma_1 \ldots \gamma_{\ell } ) \beta}
    + C_{\beta  (\gamma_1 \ldots \gamma_{\ell }) \alpha })
  }_{=:\widehat A_{\alpha\beta \gamma_1 \ldots \gamma_{\ell }}} +
B_{\alpha\beta \gamma_1 \ldots \gamma_{\ell }}
 \;,
\end{equation}
where, for reasons that will become apparent shortly, we will choose the constants $a$ and $b$  to cancel the following linear combination of the  $
E_{\alpha\beta \gamma_1 \ldots \gamma_{\ell }}$ terms in
\eqref{2VI10.3}:
\begin{eqnarray} - (n+ \ell-1)
  \widehat A_{\alpha(\beta \gamma_1\ldots \gamma_\ell)}
 +   \ell \widehat  A_{( \beta  \gamma_1 \ldots \gamma_{\ell }) \alpha }
 +  E_{\alpha\beta
\gamma_1 \ldots \gamma_{\ell }}
  = 0
 \;.
  \label{2VI10.6}
\end{eqnarray}
To check that this is possible, we calculate:
\begin{eqnarray*}
 \widehat A_{\alpha(\beta \gamma_1\ldots \gamma_\ell)}
  & = & (a+b)  E_{\alpha \beta \gamma_1\ldots \gamma_\ell }
   + b  C_{(\beta  \gamma_1 \ldots \gamma_{\ell }) \alpha }
   \;,
\\
 \ell\widehat  A_{ \beta  \gamma_1 \ldots \gamma_{\ell } \alpha  }
  & = &
     a \ell C_{\beta \gamma_1 \ldots \gamma_{\ell }\alpha}
    + b \ell( C_{\beta(\gamma_2  \ldots \gamma_{\ell }\alpha )  \gamma_1}
    + C_{  \gamma_1(\gamma_2  \ldots \gamma_{\ell }\alpha )\beta})
     \;,
\\
  & = &
     a \ell C_{\beta \gamma_1 \ldots \gamma_{\ell }\alpha}
    + b \big( C_{\alpha  \beta(\gamma_2  \ldots \gamma_{\ell } ) \gamma_1}
    +  (\ell-1) C_{  \beta(\gamma_2  \ldots \gamma_{\ell } ) \gamma_1\alpha }
\\
 &&
    +  C_{\alpha \gamma_1 (\gamma_2  \ldots \gamma_{\ell } )  \beta}
    +  (\ell-1) C_{ \gamma_1 (\gamma_2  \ldots \gamma_{\ell } ) \beta\alpha }\big)
     \;,
\\
 \ell \widehat  A_{( \beta  \gamma_1 \ldots \gamma_{\ell }  )\alpha }
  & = &
     2b   E_{\beta \gamma_1 \ldots \gamma_{\ell }\alpha}
    + (a \ell +2 b  (\ell -1))C_{(  \beta  \gamma_1\gamma_2  \ldots \gamma_{\ell })\alpha}
 \;,
\end{eqnarray*}
We thus find that \eqref{2VI10.6} is equivalent to
\begin{eqnarray}
[- (n+ \ell-1)(a+b) +2b +1] E_{\alpha\beta
\gamma_1 \ldots \gamma_{\ell }} +
\underbrace{[a \ell - ( n + 1-\ell) b]}
 C _{( \beta  \gamma_1 \ldots \gamma_{\ell }) \alpha }
  = 0
 \;.
 \qquad
  \label{2VI10.8}
\end{eqnarray}
We choose $a$ to make the underbraced term vanish,
\begin{eqnarray*}
 & \ell a = (n+1-\ell)b
 \;, &
\end{eqnarray*}
and then determine $b$ by requiring the vanishing of
\eqref{2VI10.8}:
\begin{eqnarray*}
 &  (n-1)(n+\ell +1)  b = \ell
 \;. &
\end{eqnarray*}
{Therefore the coefficients are
\begin{equation*}
a = \frac{n+1-\ell}{(n-1)(n+1+\ell)} \;,
\qquad
b = \frac{\ell}{(n-1)(n+1+\ell)} \;.
\end{equation*}
}

 Inserting \eq{2VI0.4}  in
\eqref{2VI10.2}, $\Rellp_{\alpha\beta}x^\beta$ takes now the
form
%
\begin{eqnarray}
 \nonumber
 &&
\overset{(\ell)}P_\alpha \eta_{ \beta\gamma} \myregular^\beta\myregular^\gamma|_\ell
\\
 \nonumber
 &&
 +
 \bigg(-\frac 12 \Big( 2(n+2\ell+1)
  B_{\alpha\beta \gamma_1\ldots \gamma_\ell}
 +
    \ell  (\ell-1)
 B_{\alpha\beta}{}^\mu {}_{\mu (\gamma_1\ldots \gamma_{\ell-2}}\eta_{ \gamma_{\ell-1} \gamma_{\ell })}
 \Big)
\\
 &&
  \,\,\,\,\,\,
  +
{
  \frac 12 \ell (\ell-1) B^\mu{}_{ \alpha \beta \mu (\gamma_1 \ldots \gamma_{\ell-2}}
 \eta_{\gamma_{\ell-1}\gamma_\ell)}
 + \frac 12 \ell (\ell-1) B^\mu{}_{  \beta\alpha \mu (\gamma_1 \ldots \gamma_{\ell-2}}
 \eta_{\gamma_{\ell-1}\gamma_\ell)}
 +  \ell  B_{ \alpha (\gamma_1 \ldots \gamma_{\ell }) \beta }
 }
 \nonumber
\\
 &&
  \,\,\,\,\,\,
 +   \ell  B_{ \beta (\gamma_1 \ldots \gamma_{\ell }) \alpha }
 +   \ell  B^\mu{}_{ \alpha \mu  (\gamma_1 \ldots \gamma_{\ell-1}}
 \eta_{ \gamma_\ell)\beta}
 +   \ell  B^\mu{}_{ \beta\mu  (\gamma_1 \ldots \gamma_{\ell-1}}
 \eta_{ \gamma_\ell) \alpha}
 + 2    B_{\beta \alpha \gamma_1 \ldots \gamma_{\ell-1} \gamma_\ell}
 \nonumber
\\
&&
  \,\,\,\,\,\,
   - \frac 12  \ell (\ell-1) B^\mu {}_{\mu  \beta  \alpha (\gamma_1 \ldots \gamma_{\ell-2}}
 \eta_{\gamma_{\ell-1}\gamma_\ell)}
 - \ell  B^\mu {}_{  \mu \beta  (\gamma_1 \ldots \gamma_{\ell-1}}
 \eta_{ \gamma_\ell)\alpha}
 \nonumber
\\
 &&
  \,\,\,\,\,\,
 - \ell  B^\mu {}_{\mu  \alpha (\gamma_1 \ldots \gamma_{\ell-1}}
 \eta_{ \gamma_\ell)\beta}
 -    B^\mu{}_{ \mu \gamma_1 \ldots  \gamma_\ell}
 \eta_{\beta \alpha }
 +  \eta_{\alpha(\beta}\checkC _{
\gamma_1 \ldots \gamma_{\ell })}
 +  \eta_{(\beta\gamma_1}\check C_{
\gamma_2 \ldots \gamma_{\ell })\alpha}
 \bigg)
 x^\beta x^{\gamma_1}\cdots x^{\gamma_\ell}
\nonumber
\\
 &&
 +    O(|x|^{\ell+2})
 \;,
  \label{2VI10.10}
\end{eqnarray}
for some tensors $\checkC _{ \gamma_1 \ldots \gamma_{\ell }}$ and
$\check C_{ \gamma_1 \ldots \gamma_{\ell -1 }\alpha}$.
, and
where we have denoted by $\overset{(\ell)}P_\alpha \eta_{ \beta\gamma} \myregular^\beta\myregular^\gamma|_\ell
$ the polynomial of order $\ell $ in the Taylor series of
$\overset{(\ell)}P_\alpha \eta_{ \beta\gamma} \myregular^\beta\myregular^\gamma $.
Without loss of generality we can
assume that  $\checkC _{ \gamma_1 \ldots \gamma_{\ell }}$ is
completely symmetric.

Many terms in
\eqref{2VI10.10} are proportional to $\eta_{\mu\nu}x^\mu
x^\nu$, and thus of the desired form. However, in the
homogeneous part of \eqref{2VI10.10} of order $\ell+1$ there
remain some terms proportional to $x_\alpha:=\eta_{\alpha\beta}x^\beta$ which are
\emph{not} multiplied by a factor $\eta_{\mu\nu}x^\mu x^\nu$,
and which need to be set to zero. We start by removing from \eq{2VI10.10} those terms which  obviously vanish on the light-cone; what remains is
\begin{eqnarray}
 \nonumber
 &&
\!\!\!\!\!\!
 \bigg(- (n+2\ell+1)
  B_{\alpha\beta \gamma_1\ldots \gamma_\ell}
 +  \ell  B_{ \alpha (\gamma_1 \ldots \gamma_{\ell }) \beta } +   \ell  B_{ \beta (\gamma_1 \ldots \gamma_{\ell }) \alpha }
 \nonumber
\\
 &&
 +   \ell  B^\mu{}_{ \beta\mu  (\gamma_1 \ldots \gamma_{\ell-1}}
 \eta_{ \gamma_\ell) \alpha}
 + 2    B_{\beta \alpha \gamma_1 \ldots \gamma_{\ell-1} \gamma_\ell}
  - \ell  B^\mu {}_{  \mu \beta  (\gamma_1 \ldots \gamma_{\ell-1}}
 \eta_{ \gamma_\ell)\alpha}
 \nonumber
\\
 && -    B^\mu{}_{ \mu \gamma_1 \ldots  \gamma_\ell}
 \eta_{\beta \alpha }
 +  \eta_{\alpha(\beta}\checkC _{
\gamma_1 \ldots \gamma_{\ell })}
 \bigg)
 x^\beta x^{\gamma_1}\cdots x^{\gamma_\ell}
 +  O(|x|^{\ell+2})
 \;.
  \label{2VI10.10x}
\end{eqnarray}
To continue, the tensor $ B_{\alpha\beta \gamma_1 \ldots
\gamma_{\ell }}$ in \eqref{2VI0.4} is taken of the form
\begin{equation}
 \label{2VI0.9}
B_{\alpha\beta \gamma_1 \ldots \gamma_{\ell }}=
 \eta_{\alpha\beta}B_{ \gamma_1 \ldots \gamma_{\ell }}
 \;,
\end{equation}
where $ B_{ \gamma_1 \ldots \gamma_{\ell }}$ is symmetric in
all indices.
The formula \eq{2VI10.10x} becomes, up to  terms which vanish on the light-cone,
\begin{eqnarray}
 \nonumber
 &&
  \bigg(- (n+2\ell-1)
   \eta_{\alpha\beta}B_{ \gamma_1\ldots \gamma_\ell}
 +  \ell  \eta_{ \alpha (\gamma_1} B_{ \ldots \gamma_{\ell }) \beta } +   \ell  B _{ \beta   (\gamma_1 \ldots \gamma_{\ell-1}}
 \eta_{ \gamma_\ell) \alpha}
 \nonumber
\\
 &&
  \,\,\,\,\,\,
  - \ell (n-1)  B _{ \beta  (\gamma_1 \ldots \gamma_{\ell-1}}
 \eta_{ \gamma_\ell)\alpha}  -    (n-1)B _{  \gamma_1 \ldots  \gamma_\ell}
 \eta_{\beta \alpha }
 +  \eta_{\alpha(\beta}\checkC _{
\gamma_1 \ldots \gamma_{\ell })}
 \bigg)
 x^\beta x^{\gamma_1}\cdots x^{\gamma_\ell}
\nonumber
\\
&& +  O(|x|^{\ell+2})
 \;.
  \label{2VI10.10x2}
\end{eqnarray}
Equivalently,
\begin{eqnarray}
 &&
\!\!\!\!\!
  \eta_{\alpha(\beta}\bigg(- (n+1) (\ell+2)
    B_{ \gamma_1\ldots \gamma_\ell)}
 +  \checkC _{
\gamma_1 \ldots \gamma_{\ell })}
 \bigg)
 x^\beta x^{\gamma_1}\cdots x^{\gamma_\ell}
+  O(|x|^{\ell+2})
 \;.
 \phantom{xxxxxx}
  \label{2VI10.10x3}
\end{eqnarray}
Setting
$$ B_{ \gamma_1 \ldots \gamma_{\ell }}=   \frac{1}{ (n+1) (\ell+2)}   \checkC _{
\gamma_1 \ldots \gamma_{\ell }}
 \;,
$$
the polynomial in \eq{2VI10.10x3} vanishes. This finishes the induction, and proves the result for all
$\ell \in \mathbf N$.

When $\ell= \infty$, the result is obtained by Borel-summing (see Lemma~\ref{LBorel})
the sequence of corrections $\dgell$ constructed above.
\hfill \qed
%

%
%

\bigskip

For the purposes of Theorem~\ref{T12II.1} below it is convenient to have the conclusion
of Lemma~\ref{3IV10.1} in coordinates which are harmonic for
the metric $\widehat g$. Note that the transition to such coordinates will \emph{not}  change $\tilde g$, but will in general change the remaining metric functions on $C_O$:

\begin{lemma}
 \label{L3VI0.1}
Under the hypotheses of Lemma~\ref{3IV10.1}, for any $\ell \in\N \cup \{\infty\}$ there exists a
smooth metric $\widehat g$  defined for $|\myw|$ small enough,
such that the tensor field $\tilde {\widehat g}=\hg_{AB}|_{C_O}dx^A dx^B$ induced by $\widehat g$  on $C_O$ coincides with $\tilde g$,
 such that \eq{4IV10.4b} holds
for small $|\myw|$, and the coordinates in which  \eq{4IV10.4b} holds can be chosen to be harmonic for the
metric $\widehat g$, coinciding with the original ones on the
light-cone. 
\end{lemma}

\noindent\proof
We define
$\xell^\mu$ as being normal-wave coordinates for a metric
$\gell$ defined using a modification, explained below, of the proof of
Lemma~\ref{3IV10.1}: by definition,  these are coordinates  which satisfy the wave equation
in the metric $\gell$, with $\xell^\mu$ coinciding with the
original normal coordinates $x^\mu$ on the light-cone.

Although some components of the metric tensor on $C_O$ will change when passing to the new coordinates, the $AB$ components will not. We need to marginally modify the construction of Lemma~\ref{3IV10.1} so that the introduction of
harmonic coordinates does not affect the remaining conclusions of
that Lemma,  as follows.

We start with an observation: Suppose that a function  $f+\delta f$ solves the wave equation
for a  metric $h$, given any other metric $g$ we then have
\begin{eqnarray}
 \nonumber
 0 &= &  \Box_h (f+\delta f) = h^{\mu\nu} \partial_\mu \partial_\nu  (f+\delta f)
  -  h^{\mu\nu} \Gamma(h)^\lambda{}_{\mu\nu}\, \partial_\lambda (f+\delta f)
\\ &= & (h^{\mu\nu}- g^{\mu\nu} ) \partial_\mu \partial_\nu  (f+\delta f)
 +  (g^{\mu\nu}-  h^{\mu\nu}) \Gamma(h)^\lambda{}_{\mu\nu}  \, \partial_\lambda(f+\delta f)
 \nonumber
\\
 &&
 +   g^{\mu\nu}  (\Gamma(g)^\lambda{}_{\mu\nu}-  \Gamma(h)^\lambda{}_{\mu\nu}) \, \partial_\lambda(f+\delta f)
 + \underbrace{\Box_g (f+\delta f)}_{=\Box_g  \delta f  \ \mbox{\scriptsize if $\Box_g f = 0$}}
 \;.
 \label{3VI0.1}
\end{eqnarray}

We consider \eqref{3VI0.1} with $f=x^\mu$, where $x^\mu$ denotes  normal
coordinates for the metric $g$,
and with $h:=\gellz\equiv g$, $\delta f =
\xellz^\mu - x^\mu$.
We then have $\partial (f+\delta f)  =
O(1)$, $\partial \partial(f+\delta f)  = O(1)$, $g^{\mu\nu} -
h^{\mu\nu}=O(|x|^2)$, $\Gamma (g)^\lambda{}_{\mu\nu}  =
O(|x|)$, $\Gamma (h)^\lambda{}_{\mu\nu}  = O(|x|)$, $\Box_g f =
O(|x|)$, and so \eqref{3VI0.1} implies
$$
 \Box_g (\xellz^\mu - x^\mu ) = O(|x|)
 \;.
$$
Proposition~\ref{P3IV10.2x} gives
\begin{equation}
 \label{3VI10.2}
  \xellz^\mu - x^\mu = O(|x|^3)
   \;.
\end{equation}
{}From the tensorial transformation law of the Ricci tensor, we
conclude that after the coordinate change $x^\mu \to \xellz^\mu$,
the equation
$$
\Rellz_{\alpha\beta}(O) =0
$$
will still hold in the new coordinates. Then, in the proof of
Lemma~\ref{3IV10.1} we make this coordinate change after having
constructed the metric $\gellz$ there. The construction
of the metric $\gello$ in that proof is thus done using the
coordinates $ \xellz^\mu $.

To continue, we write
$$
 \xellp^\mu=\xell^\mu+\dxell^\mu
 \;,
$$
where the notation anticipates the fact, which we are about to
prove, that the coordinates $\xellp^\mu$  differ from the
coordinates $\xell^\mu$ by terms which are $O(|x|^{\ell+3})$.
We consider \eqref{3VI0.1} with $f=\xell^\mu$,   $g=\gell$,
 $h=\gellp$, and $\delta f = \dxell ^\mu $. We again
have $\partial (f+\delta f)  = O(1)$, $\partial
\partial(f+\delta f) = O(1)$, $\Gamma (g)^\lambda{}_{\mu\nu}  =
O(|x| )$, $\Gamma (h)^\lambda{}_{\mu\nu}  = O(|x|)$, but now
$g^{\mu\nu} - h^{\mu\nu}=O(|x|^{\ell+2})$,  $\Gamma
(g)^\lambda{}_{\mu\nu}-\Gamma (h)^\lambda{}_{\mu\nu}  =
O(|x|^{\ell+1})$,  and $\Box_g f = 0$. It then follows from
\eqref{3VI0.1} that
$$
 \Box_g \dxell^\mu = O(|x|^{\ell+1})
 \;,
$$
and Proposition~\ref{P3IV10.2x} gives
\begin{equation}
 \label{3VI10.2a}
  \dxell ^\mu  = O(|x|^{\ell+3})
   \;,
\end{equation}
as anticipated by the notation.

This shows that, in the proof of Lemma~\ref{3IV10.1}, after
having constructed the metric $\gellp$, a
coordinate change
$$
 \xell\to \xellp
 \;,
$$
will preserve \eqref{4IV10.4b} (with  $x^\mu$ there equal to
$\xell^\mu$), {and for $\ell<\infty$ the proof is
completed.

If $\ell=\infty$, the construction above provides
a sequence of Taylor coefficients of the
metric which are needed so that both \eqref{4IV10.4b} and the
harmonicity vector vanish to any order. Using Borel summation, we obtain a metric for which both $R_{\mu\nu}x^\mu$ and the wave-gauge vector vanish at the vertex of the light-cone to infinite order along $C_O$. Denoting by $y^\mu$ the normal-wave coordinates for this metric, by Proposition~\ref{P3IV10.2x} we have
$$
 y^\mu - x^\mu = O(|x|^\infty)
 \;.
$$
Transforming the metric to the $y$--cordinates, the result follows.
\hfill\qed

\section{The remaining constraints: the $(\kappa,\tilde g)$ scheme}
 \label{S3V12.1}

In this section we consider the scheme of~\cite{CCM2}, where one seeks a metric which realizes the initial data $(\kappa,\tilde g)$ satisfying the first constraint equation \eq{19V10.3}. We further assume that $\tilde g$ is induced on $C_O$ by a smooth metric $C$. The analysis of Section~\ref{s10II12.1} shows how the unconstrained scheme, where $\kappa$ and the conformal class $[\tilde g]$ are prescribed, is reduced to the current one, by rescaling $C$ by a conformal factor, and calling again $C$ the resulting metric.

Let $\check C$ be the metric obtained by applying
Lemma~\ref{L3VI0.1} of Section~\ref{sS3VI0.2} to the metric $C$, so that the Ricci tensor $\check R_{\mu\nu}$ of $\checkC _{\mu\nu}$ satisfies
\bel{5V12.1}
 \underline{\check R_{\alpha\mu}}\check y^\alpha |_{C_O}= O_\ell(r^\ell)
 \;,
\ee
for any $\ell$  when  $C$ is smooth.
  This equation holds in coordinates near $O$, which we denote by  $\check \regular^\mu$,
such that $\check \regular^\mu = \regular^\mu$ on the light-cone and such
that
\bel{5V12.2}
 \Box_{\check C} \check \regular^\mu =0
  \;.
\ee
The symbols $\underline{\check C_{\mu\nu}}$ will refer to the
coefficients of the metric $\check C$ in these coordinates.
Then the coordinates $\check x^\mu$, constructed as in
\eqref{19XI.3} using the $\check \regular^\mu$'s instead of the
$\regular^\mu$'s, coincide on $C_O$ with the $x^\mu$'s. The tensor
field $\overline C_{AB} dx^A dx^B$ is intrinsic to $C_O$,
and thus coincides with $\overline {\check C}_{AB}d\check x^A d\check x^B$.
Hence, in the checked coordinates $\check x^\mu$ we still have
$$
 \check C_{AB}(\check r=r,\check x^A = x^A) = C_{AB}( r,  x^A)=: g_{AB}( r, x^A)
 \;.
$$

Let $\check H^\mu$ be the wave-map gauge vector associated with the metric $\check C$,
\bel{17VIII12.4}
 \check H^\mu := \underbrace{\check C^{\alpha\beta} ( \check \Gamma ^\mu_{\alpha\beta}}_{=:\check \Gamma^\mu}-\hat\Gamma ^\mu_{\alpha\beta}) =: \check \Gamma^\mu - \check W^\mu
 \;,
\ee
where the $\hat \Gamma ^\mu_{\alpha\beta}$'s are the Christoffel symbols of the flat metric
$$
 \hat g\equiv \eta=
-(d\check y^0)^2+  (d\check y^1)^2+\ldots+  (d\check y^n)^2 = -d\check u^2+ 2 d\check u d\check r + \check r^2 s_{AB} d\check x^A d\check x^B
 \;.
$$
It follows from \eq{5V12.2} that all the components $\underline{\check H}^\mu$ vanish, hence we have $\check H^\mu=0$ in any coordinates.

Summarising,
\bel{5V12.3}
 \overline{\check C}_{AB}=\overline C_{AB}=  \overline g_{AB}
  \ \mbox{ at $r=\check r$, $x^A = \check x^A$, and } \ \overline {\check H} ^\mu =0
  \;.
\ee
Let us denote by $\check \tau$, $\check \sigma$, etc., the fields $\tau$ and $\sigma$ associated with the metric $\check C$, e.g.
\bel{5V12.4}
 \check \chi_{AB} := \frac 12  \partial_{\widehat r} \check C_{AB}
 \;.
\ee
From \eq{5V12.3} we find in particular
\bel{5V12.3asdf}
  {\check \sigma}_{AB}= \sigma_{AB}
  \ \mbox{and $\check \tau= \tau$ at $r=\check r$, $x^A = \check x^A$.}
\ee
Set
$$
\check \kappa := \overline{\check\Gamma}^{1}_{11}
\;.
$$
Let $\ell^\mu = \check x^\mu/\check r$.
 From \cite[Equation~(6.11)]{CCM2} we have
\begin{eqnarray}
 \nonumber
O_\ell(\check r^{\ell-1})& = &
\overline{\check R}_{\mu \nu }\ell^\mu \ell^\nu
= -\partial_{1}
\check\tau
+ \overline{\check\Gamma}^{1}_{11}\tau -\check \chi_{A}{}^{B}\check\chi_{B}{}^{A}
\\
 &= &  -\partial_{1}\tau
+ \check \kappa \tau - \frac{\tau^{2}}{n-1}
- |\sigma|^{2}
 \;.
 \label{R11_13x}
\end{eqnarray}
Keeping in mind the equation satisfied by $\tau$,
\begin{equation}
\partial_{1}\tau -\kappa \tau
+ \frac{\tau^{2}}{n-1}
+ |\sigma|^{2} 
=0
\;,
 \label{19V10.3a}
\end{equation}
and using the fact that $\tau$ behaves as $(n-1)/r$ for small $r$ we conclude, at $r=\check r$, that
\bel{5VI12.5}
 \tau(\check \kappa - \kappa) = O_\ell(r^{\ell-1})
 \quad
 \Longrightarrow
 \check \kappa - \kappa = O_\ell(r^{\ell})
 \;.
\ee
To continue, recall the identities~\cite[Appendix~A]{CCM2}
\beal{27VIII12.1}
 \overline{\check \Gamma}{}^1 _ {11}
&=&
\check \nu^{0}\partial_{1} \check \nu_{0}
- \frac{1}{2}\check \nu^{0}\overline{\partial_{0} \check g_{11}}
\; ,
\\
 \label{27VIII12.2}
\check \nu_{0}\overline{\check \Gamma}{}^{0}
&=&
  \check \nu^{0}  \overline{\partial_{0} \check g_{11}}
-\frac{1}{2}  \overline{\check g}^{AB} \partial_{1}  \overline{\check g}_{AB}
=
  \check \nu^{0}  \overline{\partial_{0} \check g_{11}}
- \check \tau
\; ,%
 \\
\overline{\check W}{}^{0} &
= &
-   \check r \overline{\check g}^{AB}s_{AB}
\;,
\eeal{27VIII12.3}
hence, since $\check H^\mu \equiv \check \Gamma^\mu - \check W^\mu=0$,
\bean
\check \kappa  & \equiv  &\overline{\check \Gamma}^{1}_{11}=
\check \nu^{0}\partial_{1} \check \nu_{0}
- \frac{1}{2}\big(
\check \nu_{0}\overline{\check \Gamma}^{0}+\check \tau\big)
\\
 &= &
\check \nu^{0}\partial_{1} \check \nu_{0}
- \frac{1}{2}\big( -   \check r \check \nu_{0} \overline{\check g}^{AB}s_{AB}+\check \tau\big)
\; .
\eeal{23II12.10CE}
Keeping in mind that $\check \nu^0 = 1/\check \nu_0$ we obtain
\beal{23II12.10CE2}
 \partial_{1} \check \nu_{0}
&=&\left(
\check \kappa
+ \frac{1}{2}\big( -   (\check \nu^0 )^2\check r \overline{\check g}^{AB}s_{AB}+\check \tau\big) \right) \check \nu_ {0}
\; ,
\eea
equivalently
\bean
 \partial_{1} \check \nu^{0}
&=& -\left(
\check \kappa
+ \frac{ \check \tau}{2} \right) \check \nu^{0}+ \frac{1}{2}  \check r \overline{\check g}^{AB}s_{AB}
\\
&=& -\left(
   \kappa
+ \frac{   \tau}{2} +  O_\ell(r^{\ell}) \right) \check \nu^{0}+ \frac{1}{2}  r \overline{  g}^{AB}s_{AB}
\; .
\eeal{23II12.10CE5}
Comparing with the equation satisfied by $\nu^0$,
\begin{equation}
 \partial_1\nu^{0}=-\left(\frac{\tau}{2}+\kappa\right)\nu^{0}+ \frac{1}{2}\overline{g}{}^{AB}rs_{AB}
  \;,
 \label{5VI.1311IIEx}
\end{equation}
and using the fact that $\check \nu^0$ is smooth, hence $\check\nu^0=O_\ell(1)$ for any $\ell$,
we find
\begin{equation}
 \partial_1(\nu^{0}- \check \nu^{0})=-\left(\frac{\tau}{2}+\kappa\right)(\nu^{0}- \check \nu^{0})+  O_\ell(r^{\ell})
  \;.
 \label{5VI.1311IIExy}
\end{equation}
Integrating, we conclude that
\begin{equation}
  \nu^{0}= \check \nu^{0}+ O_\ell(r^{\ell})
  \quad
  \Longleftrightarrow
  \quad
  \nu_{0}= \check \nu_{0}+ O_\ell(r^{\ell})
  \;.
 \label{5VI.1311IIEx2}
\end{equation}

\subsection{Integration of the second constraint}
 \label{ss12II.4}

With Minkowski target  the vacuum wave-map
gauge $\mathcal{C}_{A}$ constraint reduces to~\cite{CCM2}
\begin{equation}
\mathcal{C}_{A} \equiv -\frac{1}{2}(\partial_{1}\xi_{A}+\tau\xi_{A})
+ \tilde{\nabla}_{B}\chi_{A}{}^{B}
- \partial_{A} \tau
=0
\; ,
\label{CAfinal2.2x}
\end{equation}
where $\tilde{\nabla}$ is the covariant derivative operator of the metric $\overline g_{AB}\mathrm{d}x^A \mathrm{d}x^B$, and
where $\xi_A$ is defined as
\begin{equation}
\xi_A =
- 2 \nu^{0}\partial_{1}\nu_{A} + 4 \nu^{0}\nu_{B}\chi_{A}{}^{B}
+  f_A
 \; ,
 \label{xiA.1}
\end{equation}
with~\cite[Section~8.1]{CCM2}
\begin{equation}
f_{A}=
-\left(r\overline{g}{}^{CD}s_{CD}+\frac{2\nu^{0}}{r}\right)\nu_{A}
+ \overline{g}{}_{AB}\overline{g}{}^{CD}(S_{CD}^{B}-\tilde{\Gamma}_{CD}^{B})
\; .
\end{equation}
and where the $ \tilde \Gamma{}^B_{CD}$'s are the Christoffel symbols of  the metric $\overline g_{AB}\mathrm{d}x^A \mathrm{d}x^B$.
On the other hand, for the metric ${\check{C}}$ we have  the identity
\begin{equation}
 -\frac{1}{2}(\partial_{1}{\check \xi}_{A}+\check \tau{\check \xi}_{A})
+ \tilde{\nabla}_{B}\check \chi_{A}{}^{B}
-
\partial_{A}\check \tau
 = \frac{\partial \regular^i}{\partial x^A} \overline{\underline{{\check R}_{i\nu} \ell^\nu}} = O_\ell(\check r^{\ell-1})
\; ,
\label{CAfinal2.3x}
\end{equation}
where $\check \xi_A$ is
\begin{eqnarray}
\check \xi_A &=&
- 2 \check   \nu^{0}\partial_{1}\check \nu_{A} + 4\check  \nu^{0}\check \nu_{B}\check\chi_{A}{}^{B}
 +
  \check f_{A}
  \nonumber
\\
 &=&
 - 2 \nu^{0}\partial_{1}\check\nu_{A} + 4 \nu^{0}\check\nu_{B}\chi_{A}{}^{B}
 +
  \check f_{A}
  + O_\ell(\check r^\ell)
 \; .
\label{xiA.1abc}
\end{eqnarray}
In the second line above we have used the calculations in~\cite{CCM3}, which show that
$$
 \check \nu_A =O_\ell(\check r^{3})
 \;.
$$
Further,
\begin{eqnarray}
 \nonumber
\check f_{A} & = &
-\left(\check r\overline{\check g}^{CD}s_{CD}+\frac{2\check \nu^{0}}{\check r}\right)\check \nu_{A}
+ \overline{\check g}_{AB}\overline{\check g}^{CD}(S_{CD}^{B}-\tilde{\check \Gamma}_{CD}^{B})
\\
&= &
-\left(  r\overline{  g}^{CD}s_{CD}+\frac{2  \nu^{0}}{  r} + O_\ell(r^{\ell-1}) \right)\check \nu_{A}
\nonumber
\\
 &&
+ \overline{  g}_{AB}\overline{  g}^{CD}(S_{CD}^{B}-\tilde{ \Gamma}_{CD}^{B})
 \;,%
\end{eqnarray}
at $r=\check r$.

Set
$$
 \delta \nu_A:= \nu_A - \check \nu_A
 \;, \quad
 \delta \xi_A:= \xi_A - \check \xi_A
 \;.
$$
Subtracting (\ref{CAfinal2.2x}) from (\ref{CAfinal2.3x}) one
obtains
\begin{equation}
-\frac{1}{2}(\partial_{1} \delta \xi_{A}+\tau \delta \xi_{A})
 = O_\ell(  r^{\ell-1})
\; .
\label{CAfinal2.4x}
\end{equation}
Integrating, one finds
\begin{equation}
  \delta \xi_{A}(r,x^A)
 = O_\ell(  r^{\ell})
  \;.
 \label{12II.6b}
\end{equation}
Subtracting (\ref{xiA.1abc}) from (\ref{xiA.1}) we obtain
\begin{equation}
 - 2 \partial_{1}\delta  \nu_{A}
 -  \big(   r\overline{  g}^{CD}s_{CD}  \nu^{0}  + \frac 2 r\big)\delta \nu_{A}
 + 4  \chi_{A}{}^{B}\delta  \nu_{B}
 = O_\ell(  r^{\ell})
 \; .
\label{xiA.1xx}
\end{equation}
Integrating again, Proposition~\ref{PODE1} in Appendix~\ref{s5IV0.1} gives
\begin{equation}
  \nu_{A}
 =  \check \nu_{A}+ O_\ell(  r^{\ell+1})
\; .
\label{CAfinal2.4y}
\end{equation}

\subsection{Integration of the third constraint}
 \label{S17II10.1}

We pass now to the ``${\mathcal C}_0$ constraint operator" of \cite{CCM2}. It arises from an identity, which for the $\whC$--metric takes the form
\begin{eqnarray}
0 &=&
(\ccheck{\nu}^0)^2 \Big[
2\partial_1^2(\overline{\ccheck{C}}_{00} - \overline{g}{}{}{}^{AB}\ccheck{\nu}_A\ccheck{\nu}_B)
-(\tau+4\overline{W}^1)\partial_1(\overline{\ccheck{C}}_{00} - \overline{g}{}{}{}^{AB}\ccheck{\nu}_A\ccheck{\nu}_B)
\nonumber \\ &&
\qquad\qquad +\Big(-\partial_1(\tau+2\overline{W}^1)+\overline{W}^1(\tau+2\overline{W}^1)\Big)
(\overline{\ccheck{C}}_{00} - \overline{g}{}{}{}^{AB}\ccheck{\nu}_A\ccheck{\nu}_B)
\Big]
\nonumber \\ &&
-2(\partial_1\overline{W}^1+\tau\overline{W}^1)
-\tilde{R}
+\frac{1}{2}\overline{g}{}{}{}^{AB}\ccheck{\xi}_A\ccheck{\xi}_B
-\overline{\ccheck{C}}{}{}^{AB}\tilde{\nabla}_A\ccheck{\xi}_B
\nonumber \\ &&
{-\overline{\ccheck S}_{11} \overline{\ccheck{C}}{}{}^{11}
-2\overline{\ccheck S}_{1A}\overline{\ccheck{C}}{}{}^{1A}}
-2\overline{\ccheck S}_{01}\overline{\ccheck{C}}{}{}^{01}
 \;,
  \label{25XII.2}
\end{eqnarray}
where $\check S$ is the Einstein tensor of the metric $\check C$;
here, for simplicity, we have omitted to put hats on those fields which coincide with their unhatted equivalents, e.g. $\widehat \tau=\tau$, etc. For the vacuum metric
$g_{\mu\nu}$ that we seek to construct, this provides instead a
constraint-type equation for $\overline g _{00}$:
\begin{eqnarray}
0 &=&
(\nu^0)^2 \Big[
2\partial_1^2(\overline{g}{}_{00} - \overline{g}{}{}{}^{AB}\nu_A\nu_B)
-(\tau+4\overline{W}^1)\partial_1(\overline{g}{}_{00} - \overline{g}{}{}{}^{AB}\nu_A\nu_B)
\nonumber \\ &&
\qquad\qquad +\Big(-\partial_1(\tau+2\overline{W}^1)+\overline{W}^1(\tau+2\overline{W}^1)\Big)
(\overline{g}{}_{00} - \overline{g}{}{}{}^{AB}\nu_A\nu_B)
\Big]
\nonumber \\ &&
-2(\partial_1\overline{W}^1+\tau\overline{W}^1)
-\tilde{R}
+\frac{1}{2}\overline{g}{}{}{}^{AB}\ccheck{\xi}_A\ccheck{\xi}_B
-\overline{g}{}{}{}^{AB}\tilde{\nabla}_A\ccheck{\xi}_B
 \;.
  \label{25XII.1}
\end{eqnarray}
Subtracting (\ref{25XII.2}) from (\ref{25XII.1}) we obtain an
ODE for $\overline{ \ccheck{C}_{00}- g_{00}}$ which, as before,
leads to
$$
 \overline{g}{}_{00}  =\overline{ \ccheck{C}}_{00}+   O_\ell(  r^{\ell }) \;.
$$
To establish this, the reader might find it convenient to argue in two steps, by first considering the first-order ODE  satisfied by the difference of $\partial_1(\overline{g}{}_{00} - \overline{g}{}{}{}^{AB}\nu_A\nu_B)$ and $\partial_1(\overline{C}{}_{00} - \overline{g}{}{}{}^{AB}\nu_A\nu_B)$.

\subsection{End of the proof}
 \label{S25VIII12.1}

Let $C_{\mu\nu}$ be a smooth metric and let $\kappa$ be a  function on $C_O$ such that $\kappa/r$ extends to a smooth function on space-time.

{}From  what has been said, there exist smooth space-time
functions  $\delta\ccheck{ C}_{0A}$, $\delta\ccheck{ C}_{01}$
and $\delta\ccheck{ C}_{00}$   vanishing to infinite order at
the origin such that
$$
 \overline{\delta\ccheck{ C}_{0A} }= -\overline{\ccheck{C}_{0A}}    +{\nu_{A}}\;, \qquad
 \overline{\delta\ccheck{ C}_{00}} =  -\overline{\ccheck{C}_{00}}    +\overline{g}{_{00}}\qquad
 \overline{\delta\ccheck{ C}_{01}} =  -\overline{\ccheck{C}_{01}}    +{\nu_{0}}
 \;.
$$
Then the tensor field $\delta\ccheck{ C}$ defined as
 $$
\delta\ccheck{ C}:= 2\delta\ccheck{ C}_{01} du\, dr + 2\delta\ccheck{ C}_{0A} du\, dx^A + \delta\ccheck{ C}_{00} du^2
 $$
has smooth components $\underline{\delta\ccheck{ C}_{\mu\nu}}$,
and satisfies
$$
 \overline{\delta\ccheck{ C}_{AB}} = 0 =
 \overline{\delta\ccheck{ C}_{A1}} =
 \overline{\delta\ccheck{ C}_{11}}
  \;.
$$
It follows that the tensor
$$
  \underline{{\ccheck{C}_{\mu\nu}}  +
  {\delta\ccheck{ C}_{\mu\nu}}}
$$
has smooth components, satisfies the Raychaudhuri constraint equation
\eqref{19V10.3} with prescribed function $\kappa$, as well as the remaining wave-map gauge constraint
equations. The existence theorem  of
\cite{Dossa97} shows existence of a smooth metric $g_{\mu\nu}$, defined in a neighborhood of the vertex $O$, which satisfies the vacuum Einstein equations to the future of $O$, such that
$$
 \underline{g_{\mu\nu}}|_{C_O}= \underline{{\ccheck{C}_{\mu\nu}}  +
  {\delta\ccheck{ C}_{\mu\nu}}}
  \;.
$$
It then follows form the analysis in~\cite{CCM2} that $H^\mu\equiv 0$ (compare the argument at the end of Section~\ref{S28VII12.1}), and that $g_{\mu\nu}$ solves the Einstein vacuum equations to the future of $O$, with
$$
\overline{\Gamma^1_{11}}=\kappa
 \;.
$$

We have therefore proved:

\begin{theorem}
 \label{T12II.1}
Consider a pair
$(\kappa,\tilde g)$, where $\tilde g$ is a symmetric tensor field induced by a smooth Lorentzian metric  $C$ on its null cone $C_O$ with vertex  at $O$, and where $r\kappa$ is the restriction to $C_O$ of a smooth function on space-time vanishing to second order at $O$.
Suppose moreover that
$(\kappa,\tilde g)$ satisfy the Raychadhuri equation
\begin{equation}
\partial_{1}\tau -\kappa \tau
+ \frac{\tau^{2}}{n-1}
+ |\sigma|^{2} 
=0
\;,
 \label{19V10.3xy}
\end{equation}
where $\tau$ is the divergence of $C_O$ and $\sigma$ its shear.  Then
there exists a smooth  metric $g$, defined in   neighborhood of $O$ and solving  the vacuum Einstein
equations  in $J^+(O)$, such that $C_O$ is the light-cone of $g$,   $\tilde g$
is the tensor field induced by $g$ on $C_O\setminus\{O\}$, and $\kappa$ determines parallel-transport along the generators of $C_O$: in adapted coordinates
$$
 \nabla_{\partial_r}\partial_r = \kappa \partial_r
 \;.
$$
\hfill\qed
\end{theorem}

We note that \eq{19V10.3xy} is a necessary condition for $g$ to be vacuum, so Theorem~\ref{T12II.1} is in fact an if-and-only-if statement.

\section{The $\overline g_{\mu\nu}$ scheme}
 \label{S28VII12.1}

In this section we prove Theorem~\ref{T29VOOO12/1}, namely  existence  of solutions of the vacuum Cauchy problem on the light-cone in the scheme of \cite{ChPaetz}, where all the metric functions are prescribed by restricting   a smooth metric $C$ to its light-cone.

As in our previous treatment, we use a ``generalized wave-map gauge"  with
target metric $\hat g$ being the Minkowski metric $\eta=-(dy^0)^2+(dy^1)^2
+\cdots + (dy^n)^2$.
As gravitational initial data, we choose a smooth tensor field $C$. The coordinates $y$ are chosen so that the future light-cone $C_O$ of $C$ with vertex at $O$ coincides with the Minkowskian light-cone $y^0 = |\vec y|$. We then use the metric  components $\overline{  {C_{\mu\nu}}}=C_{\mu\nu}|_{C_O}$ as initial data for $g$:
$$
\overline   {g} _{\mu\nu} : = \overline   {C}_{\mu\nu}
 \;.
$$
It follows from Lemma~\ref{3IV10.1} that there exists a metric $\check C$ such that
\bel{27VIII12.6}
\overline   {g} _{\mu\nu}  = \overline   {C}_{\mu\nu} = \overline   {\check C}_{\mu\nu}
 \;,
\ee
with the Ricci tensor $\check R_{\mu\nu}$ of the metric $\check C$ satisfying the conclusions of that lemma:
for small $r\equiv |\vec y|$,
\bel{26VIII12.1}
  \check R _{\mu\nu}  = O(r^{ 2 })
 \;,
\quad
  \check R_{\mu\nu}y^\nu |_{C_O}= \Oiri
 \;.
\end{equation}

To obtain a well posed system of evolution equations for the metric $g$ we will impose a \emph{generalized wave-map gauge condition},
\begin{equation*}
 H^{\lambda}=0
   \;,
\end{equation*}
with the \emph{harmonicity vector}  $H^\mu$ defined as
\begin{equation}
H^{\lambda} :=
\underbrace{g^{\alpha\beta} \Gamma_{\alpha\beta}^{\lambda}}_{=:\Gamma^\lambda}-W^{\lambda}
\;, \enspace \mathrm{with}\enspace
W^{\lambda}:=
 \underbrace{g^{\alpha\beta} \hat{\Gamma}_{\alpha \beta}^{\lambda}}_{=:\timW^\lambda} + \What^{\lambda}
\;,
\label{WaveGauge0x}
\end{equation}
where the $  \hGamma _{\alpha\beta}^{\lambda}$'s are the Christoffel symbols of the metric $\eta\equiv \hg$. Roughly speaking, we calculate $\ol \Gamma{} ^\lambda-\ol \timW{}^\lambda$ from the initial data, and use the result as the definition of $\ol \What{}^\lambda$; this will ensure the vanishing of $\ol H ^\mu$. The details  are somewhat less straightforward, as  $\ol H{} ^\lambda-\ol \timW{}^\lambda$ involves some transverse derivatives of the metric which are not part of the initial data; this is taken care of as in~\cite{ChPaetz}.
One then needs to prove that  $\overline{\What }{}^\mu$ is the restriction to the light-cone of  a smooth vector field in space-time, and this is focus of the work here.

Recall that the vector field $\check H^\mu$ has been defined in \eq{17VIII12.4} as
\begin{equation}
\check H^{\lambda} :=
\underbrace{\check C^{\alpha\beta} (\check \Gamma _{\alpha\beta}^{\lambda}}_{=:\check \Gamma {}^\lambda}-  \hat{\Gamma}_{\alpha \beta}^{\lambda} )
\;,
\label{WaveGauge0x}
\end{equation}
where the $ \check \Gamma _{\alpha\beta}^{\lambda}$'s are the Christoffel symbols of the metric $\check C$. This is clearly a smooth vector field in space-time. We will show that the components of $\overline{\What }{}^\mu$ differ  from those of $\overline{\check H }{}^\mu$
 by  terms  which are $\Oiri $.
It easily follows from Lemma~\ref{L23I.1}, Appendix~\ref{ss8VIII0.1}, that a vector field, defined along $C_O$, with $(u,r,x^A)$-components that are $\Oiri $ extends to a smooth vector field on space-time, which will establish the desired property of $\overline{\What }{}^\mu$.

We pass now to the details of the above.
There exists a neighbourhood of $O$ on which $\tau$ has no zeros. There
we solve the first constraint by setting
\begin{eqnarray}
 \kappa &=&\frac{\partial_1\tau + \frac{1}{n-1}\tau^2 + |\sigma|^2  }{\tau}
  \;.
 \label{relation_kappa-tau26VIII}
\end{eqnarray}
The argument leading to \eq{5VI12.5} applies, and gives
\bel{5VI12.5x}
 \check \kappa - \kappa = \Oiri
 \;.
\ee
Following~\cite{ChPaetz}, we choose $\overline \What{}^ 0$ to be
\begin{eqnarray}
 \overline \What {}^0 &=& -\overline{\timW }{}^0- \nu^0(2\kappa+\tau) - 2 \partial_1\nu^0
  \;;
   \label{27VIII12.5}
\end{eqnarray}
equivalently, using the unchecked versions of \eq{27VIII12.1}-\eq{27VIII12.3},
\bel{23II12.10}
 \overline \Gamma{}^1 _ {11} = \kappa - \frac 12 \nu_0\overline H{} ^0
 \;.
\ee
The last equation is further equivalent to (compare the unchecked version of \eq{R11_13x})
\bel{23II12.2}
 \overline{R}_{11}   = - \frac 12  \nu_0\overline H{} ^0 \tau
 \;.
\ee

Comparing the definition  \eq{WaveGauge0x} of $\check H$ with \eq{27VIII12.5}, using \eq{27VIII12.6} and \eq{5VI12.5x} we find
\bel{27VIII12.7}
 \overline \What {}^0  = \overline{\check H}{}^0 + 2\nu^0(\check\kappa - \kappa) = \overline{\check H}{}^0 + \Oiri
 \;.
\ee

The next constraint equation follows from $\overline R_{2A}  =0$. We note the identity~\cite{CCM2}
\begin{eqnarray}
     (\partial_r + \tau)  \overline\Gamma{}^1_{1A}  + \tilde{\nabla}_B \sigma_A^{\phantom{A}B} - \frac{n-2}{n-1}\partial_A\tau -\partial_A \ol\Gamma{}^1_{11}
 = \overline R_{2A}
 \;,
 \label{20XII11.1}
\end{eqnarray}
where $\tilde\nabla$ is the covariant derivative associated to the Riemannian metric $g_{AB}$.

We let $\xi_A$ to be the unique solution, which vanishes at the tip of the light-cone, of the equation obtained by replacing $  \overline\Gamma{}^1_{1A}$ in \eq{20XII11.1} by $- \xi_A/2$, $\ol\Gamma{}^1_{11}$ by $\kappa$, and setting the right-hand side to zero,
\begin{eqnarray}
   - \frac{1}{2}(\partial_r + \tau)\xi_A + \tilde{\nabla}_B \sigma_A^{\phantom{A}B} - \frac{n-2}{n-1}\partial_A\tau -\partial_A \kappa
 = 0
 \;,
 \label{20XII11.1x}
\end{eqnarray}
as in \eq{CAfinal2.2x}.
We choose $\overline{\What}{}^A$ to be
\begin{eqnarray}
 \overline \What {}^A &:=&\overline g^{AB}\Big[\xi_B + 2\nu^{0} (\partial_r\nu_{B} - 2 \nu_{C}\sigma_B{}^C -\nu_{B} \tau)  -\nu_{B}(\overline \What {}^0 + \overline{\timW }{}^0 ) \Big]
  \nonumber
\\
&&   +\overline g^{CD}\tilde\Gamma^A_{CD} - \overline{\hat W}{}^A
 \;;
  \label{23II12.5}
\end{eqnarray}
equivalently
\begin{eqnarray}
   \xi_A =  -2\nu^{0}\partial_r\nu_{A} + 4\nu^{0}\nu_{B}\sigma_A{}^B + 2\nu^{0}\nu_{A}\tau + \nu_{A}(\overline \What{}^0+\overline{\timW }{}^0)
 \nonumber
\\
   + \overline g_{AB}(\overline \What{}^B + \overline{\timW }{}^B) - \overline g_{AB} \overline g^{CD} \tilde \Gamma{}^B_{CD}
 \;.
 \label{20XII11.2}
\end{eqnarray}
This has been chosen so that, using the formulae in \cite[Appendix~A and Section~9]{CCM2},
\begin{equation}
\overline{R}_{1A} =
-\frac{1}{2}(\partial_r+\tau)(\og_{AB}\overline{H}{}{}^{B}+\nu_{A}\overline{H}{}{}^{0}) +\frac{1}{2}\partial_{A}(\nu_0\overline{H}{}{}^{0})
\; .
\label{LAfinal}
\end{equation}
Moreover, one finds (cf.\ Equation (10.35) in \cite{CCM2})
\begin{equation}
 \xi_A = -2\overline\Gamma{}^1_{1A} - \overline g_{AB} \overline H{}^B -  \nu_{A} \overline H{}^0
 \;.
 \label{relation_xiA_HA}
\end{equation}

We let $\cxi _A$  be  $-2 \overline{\check\Gamma}{}^1_{1A}$. The check-equivalent of \eq{20XII11.1} reads
\begin{eqnarray}
   - \frac{1}{2}(\partial_r + \tau)\cxi_A+ \tilde{\nabla}_B \sigma_A^{\phantom{A}B} - \frac{n-2}{n-1}\partial_A\tau -\partial_A \check\kappa
 = \Oiri
 \;.
 \label{20XII11.1check}
\end{eqnarray}
Comparing with the equation \eq{20XII11.1x} defining $\xi_A$ we find
\begin{eqnarray}
   - \frac{1}{2}(\partial_r + \tau)(\xi_A - \cxi_A)
 = \Oiri
 \;.
 \label{20XII11.1check1}
\end{eqnarray}
Integration establishes that
\begin{eqnarray}
   \xi_A = \cxi_A+   \Oiri
 \;.
 \label{20XII11.1checka}
\end{eqnarray}
The field $\check H^A$, defined in \eq{WaveGauge0x} and written out in detail using \cite[Appendix~A]{CCM2}, takes the form
\begin{eqnarray}
 \overline{\check H} {}^A & =&\overline g^{AB}\Big[\cxi_B + 2\nu^{0} (\partial_r\nu_{B} - 2 \nu_{C}\sigma_B{}^C -\nu_{B} \tau)  -\nu_{B}  \overline{\check \Gamma }{}^0   \Big]
  \nonumber
\\
&&   +\overline g^{CD}\tilde\Gamma^A_{CD} - \overline{\hat W}{}^A
 \;.
  \label{23vIII12.10}
\end{eqnarray}
Comparing with \eq{23II12.5} and using \eq{27VIII12.7} and \eq{20XII11.1checka} we conclude that
\begin{eqnarray}
 \nonumber
 \overline{  \What } {}^A & =&\overline{\check H} {}^A + \overline g^{AB}\Big[\xi_B-\cxi_B    -\nu_{B}  ( \overline \What {}^0 + \overline{\timW }{}^0 -\overline{\check \Gamma }{}^0  ) \Big]
\\
  & =&\overline{\check H} {}^A +
 \Oiri
 \;.
  \label{23vIII12.11}
\end{eqnarray}

Let $S_{\mu\nu}$ denote the Einstein tensor of $g$.
We continue with the equation $\overline S_{01}   = 0$; equivalently,
$\overline g^{AB}\overline R_{AB} = 0$.  Using the identities (10.33) and a corrected version of%
\,
\footnote{On the far-right-hand side of (10.36) in \cite{CCM2}  a term $\tau \overline{g}^{11}/2$ is missing.}
\cite[Equation~(10.36)]{CCM2} we find the identity
\begin{eqnarray}
  \overline g^{AB} \overline R_{AB} & \equiv & (\partial_r + \overline \Gamma{}^1_{11} + \tau)(2\overline g^{AB} \overline\Gamma{}^1_{AB} +\tau\ol g^{11})
   \nonumber
\\
  && - 2\overline g^{AB} \overline\Gamma{}^1_{1A} \overline\Gamma{}^1_{1B}
 - 2 \overline g^{AB} \tilde\nabla_A \overline\Gamma{}^1_{1B}  + \tilde R
 \;.
  \label{28VIII12.1}
\end{eqnarray}
This motivates the equation
\begin{eqnarray}
 (\partial_r + \kappa + \tau) \zeta   + \big( \tilde\nabla_A - \frac{1}{2}\xi_A\big) \xi^A + \tilde R =  0
 \;,
 \label{constraint3}
\end{eqnarray}
with $\xi^A:=\overline g^{AB} \xi_B$, and
where the quantity $\zeta$ will be the restriction of
\begin{equation*}
  2 \big(\overline g^{AB} \overline\Gamma{}^1_{AB}+ \tau \ol g^{11}
 \big)
\end{equation*}
to $C_O$, once the final vacuum metric $g$ will have been constructed.
We integrate  (\ref{constraint3}),  viewed as a first-order ODE for $\zeta$, as
\begin{eqnarray*}
   \zeta &=& - \frac{ e^{-\int_1^r (\kappa+\tau-\frac{n-1}{\tilde r})\mathrm{d}\tilde r}}{r^{n-1}}  \int_{0}^r \tilde r^{n-1} e^{\int_1^{\tilde r} (\kappa+\tau-\frac{n-1}{\tilde{\tilde r}})\mathrm{d}\tilde{\tilde r}}\times
\\
 &&\phantom{xxxxxxxxxxxxxxxxxxx}
   \Big( \tilde R   + \overline g^{AB}\tilde{\nabla}_A \xi_B  - \frac{1}{2}\overline g^{AB}\xi_A\xi_B  \Big) \mathrm{d}\tilde r
 \\
  & = & -(n-1) r^{-1} + O(1)
 \;.
\end{eqnarray*}

We choose
\begin{equation}
\overline \What {}^1  :=   \frac 12 \zeta  - (\partial_r+\kappa + \frac 12 \tau )\bar{g}^{11}-\overline{\timW }{}^1\;;  \label{9.3x}
\end{equation}
equivalently
\begin{eqnarray}
 \zeta= 2\overline g^{AB}\overline \Gamma{}^1_{AB} + \overline g^{11}(\tau +  \nu_{0} \overline H{}^0 )- 2 \overline H{}^1
 \;.
 \label{relation_zeta_H2}
\end{eqnarray}
With the choice  (\ref{9.3x}) we have
\begin{eqnarray}
 \overline g^{AB} \overline R_{AB}
& = &  (\partial_r +\kappa +  \tau -\frac{1}{2}\overline g_{12} \overline H{}^0 ) (2\overline H{}^1  - \overline g_{12}\nu^{0}\overline H{}^0 )    -\frac{1}{2}\overline g_{12} \overline H{}^0\zeta
 \nonumber
\\
   &&
   +(\tilde\nabla_A- \xi_A - \frac{1}{2}\overline g_{AB}\overline H{}^B - \frac{1}{2}\nu_{A}\overline H{}^0)( \overline H{}^A + \nu_{C} \overline g^{AC}\overline H{}^0)
 \;.
  \phantom{xxx}
 \label{L0final}
\end{eqnarray}

Let $\check \zeta$ be the check-counterpart of $\zeta$,
\begin{equation}
 \check \zeta:= 2 \big(\overline g^{AB} \overline{ \check\Gamma}{}^1_{AB}+ \tau \nu^{0}
 \big)
 \;.
  \label{28VIII12.7}
\end{equation}
Integrating the check-version of (\ref{28VIII12.1}) we obtain
\begin{eqnarray}
 \nonumber
  \check \zeta &=& - \frac{ e^{-\int_1^r (\check \kappa+\tau-\frac{n-1}{\tilde r})\mathrm{d}\tilde r}}{r^{n-1}}  \int_{0}^r \tilde r^{n-1} e^{\int_1^{\tilde r} (\check \kappa+\tau-\frac{n-1}{\tilde{\tilde r}})\mathrm{d}\tilde{\tilde r}}\times
\\
 \nonumber
 &&\phantom{xxxxxx}
    \Big( \tilde R   + \overline g^{AB}\tilde{\nabla}_A \cxi_B  - \frac{1}{2}\overline g^{AB}\cxi_A\cxi_B
   - \ol g^{AB} \ol {\check R}_{AB} \Big) \mathrm{d}\tilde r
 \\
  & = &
   \zeta + \Oiri
 \;.
  \label{28VIII12.5}
\end{eqnarray}
From \eq{28VIII12.7} and from \cite[Appendix~A]{CCM2} we find
\bel{28VIII12.6}
 \frac 12 \check \zeta =  \ol{\check \Gamma}{}^1 + (\partial_r +\check \kappa + \frac 12 \tau)
 \ol g ^{11}
 \;.
\ee
Comparing with \eq{9.3x}, in view of \eq{5VI12.5x} and \eq{28VIII12.5}  we conclude that
\bel{28VIII12.9}
 \overline{\What}{}^1 =  \frac 12 (\zeta -\check \zeta) + \ol{\check \Gamma}{}^1 + (\check \kappa -\kappa )\bar{g}^{11}-\overline{\timW }{}^1=
 \overline{\check H} {}^1 + \Oiri
 \;.
\ee

Summarising, given the fields  $\overline g_{\mu\nu}$  on $C_O$  we have found a vector field  $\overline{\What}$ on  $C_O$ satisfying
$$
 \overline{\What}{}^\mu = \overline{\check H} {}^\mu + \Oiri
 \;.
$$
The field $ \overline{\check H} {}^\mu $ extends trivially to the smooth vector field $ {\check H} {}^\mu $, while a vector field with components which are $\Oiri$ extends to a smooth vector field in space-time by Lemma~\ref{L23I.1}. We conclude that there exists  a  smooth vector field, which we call $\What$, defined in a neighborhood  $
\mcO$ of $O$,  which coincides with $\overline\What$ on $C_O\cap \mcO$.

We apply the existence and uniqueness theorem of~\cite{DossaAHP} to the reduced Einstein equations $ R_{\alpha\beta}^{(H)}=0$, with initial data $\overline g$, where
\begin{equation}
R_{\alpha\beta}^{(H)}:=R_{\alpha\beta} -{\frac{1}{2}}(g_{\alpha\lambda
}\hat{D}_{\beta}H^{\lambda}+g_{\beta\lambda}\hat{D}_{\alpha}H^{\lambda
}) ,
\label{RicciHIdentity}
\end{equation}
with $H^\mu$ defined by  (\ref{WaveGauge0x}), where
$\hat{D}$  is  the Levi-Civita covariant derivative in the metric
$\hat{g}$.
Indeed, it follows from~\cite[page 163]{YCB:GRbook} that $R_{\alpha\beta}^{(H)}$ is a
quasi-linear, quasi-diagonal operator on $g$, tensor-valued,
depending on $\hat{g}$, of the form
\begin{equation}
R_{\alpha\beta}^{(H)}\equiv-{\frac{1}{2}}g^{\lambda\mu}
{\hat{D}}_{\lambda} {\hat{D}}_{\mu}g_{\alpha\beta}+
\hat{f}[g,{\hat{D}}g]_{\alpha\beta}\;,
\label{RicciH}
\end{equation}
where $\hat{f}[g,{\hat{D}}g]_{\alpha\beta}$  is a tensor quadratic in $\hat Dg$
with coefficients  depending upon  $g$,  $\hat{g}$,    $\What$,  $\hat D \timW$ and $\hat D \What$, which is of the right form for~\cite{DossaAHP}.

Now, the metric $g$ so obtained will solve the vacuum Einstein equations if
and only if $H^\mu$ vanishes on $C_O$. It should  be clear that $\kappa$ equals then $\ol {\Gamma}{}^1_{11}$  and   $\xi_A$ equals $-\frac 12 \ol \Gamma{}^1_{1A}$ but, to avoid ambiguities, we will justify it explicitly in what follows.

Note that at this stage a smooth metric $g$ and   smooth vector fields  $ \What^{\mu}$ and $H^\mu$
are known in a neighbourhood $\mcU$ of $O$, with $g$   satisfying the reduced Einstein equations  in $\mcU\cap J^+(O)$.

The proof of vanishing of $\ol H$ is essentially the same as the one in~\cite{CCM2}, we outline it here for completeness.

In order to prove that $\ol H{}^0 =0$ holds, we note the identity  (see
(\ref{RicciHIdentity}))
\begin{equation}
\overline{R}_{11} \equiv
\underbrace{\overline{R}_{11}^{(H)}}_{=0}+\nu_{0}\hat D_{1}\overline{H}{}^{0}
 \;.
 \label{28VIII12.15}
\end{equation}
The reader will note that this equation, as well as \eq{HApreserve1} and \eq{H0preserve_3b} below, are identical with the corresponding equations in \cite{CCM2}, even though \emph{our $H$ is not the same as the corresponding vector field in \cite{CCM2}.} This is due to the fact that our operator $\overline{R}_{11}^{(H)}$ in \eq{RicciHIdentity} is constructed using our vector field $ \Gamma^\mu - \hat W^\mu - \What{}^\mu $, while in \cite{CCM2} the vector field $\Gamma^\mu - \hat W^\mu$ is used for $H^\mu$.

Equations \eq{23II12.2} and \eq{28VIII12.15} imply that $\overline{H}{}^{0}$ satisfies a linear homogeneous
differential equation on $C_{O}$, namely,
\begin{equation}
 \label{19V.1}
 \hat D_{1}\overline{H}{}^{0}+\frac{1}{2}\tau \overline{H}{}^{0}= 0\;.
\end{equation}
As explained in \cite[Section~7.6]{CCM2}, the only bounded solution of this equation is  $\ol H {}^0 \equiv 0$.  The equality $\Gamma^1_{11}|_{C_O}=\kappa $   follows trivially now from  \eq{23II12.10},
\begin{eqnarray}
 \label{23II12.1asdf}
  H^0|_{C_O} \equiv 2\nu^{0} (\kappa- \Gamma^1_{11})
 \;.
\end{eqnarray}

To establish the vanishing of $\ol H{}_A$ we invoke the identity \cite[Equation~(9.8)]{CCM2}:
\begin{equation}
\overline{R}_{1A}\equiv
\underbrace{\overline{R}_{1A}^{(H)}}_{=0}+
\frac{1}{2}(\nu_{0}\overline{\hat{D}_{A}H^{0}}+
        \nu_{A}\overline{\hat{D}_{1}H^{0}}+
        \overline{g}_{AB}\overline{\hat{D}_{1}H^{B}})
        \;.
\label{HApreserve1}
\end{equation}
Combined with \eq{LAfinal}, and taking into account that $\ol H {}^0=0$ has already been established, this gives a radial homogeneous ODE for $\ol H {}  ^A$, with $\ol H {}^A\equiv 0$ being the only solution with the relevant asymptotic behaviour at $O$. We can now  conclude that $\xi_A = -2\overline\Gamma{}^1_{1A}  $ from \eq{relation_xiA_HA}.

Finally, we have the identity~\cite[Equation~(11.18)]{CCM2} (recall that $S_{\mu\nu}$ denotes the Einstein tensor)
\begin{equation}
\overline{S}_{01}\equiv \underbrace{\overline{S}_{01}^{(H)}}_{=0}+
\frac{1}{2}(\overline{g}_{00}\overline{\hat{D}_{1}H^{0}}+
            \nu_{A}\overline{\hat{D}_{1}H^{A}}-
            \nu_{0}\overline{\hat{D}_{A}H^{A}}).
\label{H0preserve_3b}
\end{equation}
Combining this with \eq{L0final}, one similarly concludes that $\ol H{}^1=0$; see also \cite[Section~11.3]{CCM2}.
The vanishing of  $\overline H{}^0$ and  $\overline H{}^1$, together with the identity  (\ref{relation_zeta_H2}), imply that  on $C_O$ the field $\zeta$  coincides with $2\overline g^{AB} \overline\Gamma{}^1_{AB} +  \tau \nu^{0}$.

Thus $H^\mu$ vanishes on $C_0$, and by the usual arguments (see, e.g., \cite[Theorem~3.3]{CCM2}) we have $H^\mu \equiv 0$.

This completes the proof of Theorem~\ref{T29VOOO12/1}.
\qed

\appendix

\section{On Taylor expansions}
 \label{ss8VIII0.1}

To proceed, some terminology will be needed. We say that a
function $g$ defined on a space-time neighbourhood of the
origin is $o_m(|y|^k)$ if $g$ is $C^m$ and if for $0\le \ell
\le m$ we have
$$
 \lim_{|y|\to 0}
 |y|^{\ell-k}\partial_{\mu_1} \ldots
 \partial_{\mu_\ell} g =0
 \;,
$$
where $|y|:=\sqrt{\sum_{\mu=0}^n (y^\mu)^2}$.

A similar definition will be used for functions defined in a
neighbourhood of $O$ on the light-cone $C_O$:  We parameterize
$C_O$ by  coordinates $y^i\in {\R }^n$, and  we say that
a function $g$ defined on a  neighbourhood of $O$ within $C_O$
is $o_m(r^k)$ if $g$ is a $C^m$ function   of the coordinates
$y^i$ and if for $0\le \ell \le m$ we have $\lim_{r\to 0}
r^{\ell-k}\partial_{\mu_1} \ldots
\partial_{\mu_\ell} g =0$, where $r:=\sqrt{\sum_{i=1}^n (y^i)^2}$.

 We consider a  light-cone $C_O$ which is smooth away from
its tip. The following lemma will be used repeatedly (recall that $\Theta^i= y^i/r$):

\begin{lemma}[Lemma~A.1 in \cite{ChJezierskiCIVP}]
 \label{L23I.1}
A function $\varphi$ defined on a light-cone $C_{O}$ is the
trace $\overline{f}$ on $C_{O}$ of a $C^{k}$ spacetime function
$f$ if and only if $\varphi$ admits an expansion of the form
\begin{equation}
 \label{23XI.1}
\varphi= \sum_{p=0}^{k}f_{p}r^{p}+o_k(r^{k})
 \;,
\end{equation}
with
\begin{equation}
 \label{6XI.1}
f_{p}\equiv
f_{i_{1}\ldots i_{p}}\Theta ^{i_{1}}\cdots \Theta ^{i_{p}}
+
f^{\prime}{}_{i_{1}\ldots i_{p-1}}\Theta ^{i_{1}}\cdots \Theta ^{i_{p-1}}
 \;,
\end{equation}
where $f_{i_{1}\ldots i_{p}}$ and $f_{i_{1}\ldots
i_{p-1}}^{\prime }$ are numbers.

The claim remains true with $k=\infty$ if
 \eqref{23XI.1} holds for all $k$.
  \qed
\end{lemma}

We will also need the following:

\begin{lemma}[Borel summation, see e.g.~Lemma~D.1 in \cite{ChJezierskiCIVP}]
 \label{LBorel}
 For any sequence $\{c_{i_1\ldots i_k}\}_{k\in\bf N}=\{c, c_i, c_{ij},\ldots\}$ there exists a smooth
function $f$ such that, for all $k\in \mathbf N$,
$$
  f- \sum_{p=0}^k c_{i_1\ldots i_p} y^{i_1}\cdots y^{i_p} = o_k(r^k)
   \;.
 $$
\qed
\end{lemma}

\section{ODE Lemmas}
 \label{s5IV0.1}

For $k\in \N\cup\{\infty,\omega\}$ we will say that a function $\varphi:C_O\to \R$ is $C^k$-cone differentiable if there exists a $C^k$  function on space-time $\phi$ such that $\varphi$ is the restriction to $C_O$ of $\phi$. We shall say ``cone-smooth" for $C^\infty$-cone differentiable.

We start with the following elementary result:

\begin{lemma}
 \label{L1V12.1}
Let  $k\in \N\cup\{\infty,\omega\}$, and let $\varphi$ be a $C^k$-cone differentiable function on $C_O$.
Then the integrals
\bel{1V12.2}
 \psi(r,x^A) = \int_0^r \frac{\varphi(s,x^A)}s ds
 \quad \mbox{\rm and} \quad
 \chi(r,x^A) = \frac 1 r \int_0^r  {\varphi(s,x^A)}  ds
\ee
are $C^k$-cone differentiable, assuming moreover $\varphi(0)=0$ in the case of the integral defining $\psi$.
\end{lemma}

\proof
Let, first, $k\in \N$. By Lemma~\ref{L23I.1} we have
\begin{equation}
 \label{23XI.1x}
\varphi= \sum_{p=0}^{k}f_{p}r^{p}+o_k(r^{k})
 \;,
\end{equation}
where the coefficients $f_p$ are of the form \eq{6XI.1}. Inserting into \eq{1V12.2} we find
\bel{1V12.3}
 \psi(r,x^A) = \sum_{p=1}^{k}\frac{f_{p} r^{p }}{p }+o_k(r^{k})
 \;,
 \qquad
 \chi(r,x^A) = \sum_{p=0}^{k}\frac{f_{p} r^{p }}{p+1 }+o_k(r^{k})
 \;,
\ee
and the result follows from Lemma~\ref{L23I.1}.

The case $k=\infty$ is established in a similar way using Borel summation.

The case $k=\omega$ is the contents of~\cite[Lemma~6.5]{CCM4}.
\qed

\medskip

We will need the following result about systems of Fuchsian ODEs:
\begin{lemma}
 \label{LODE1}
 Let $r_0>0$, $k\in\N \cup\{\infty\}$, $N\in\N $, $0>a \in \mathbf R$, $ \psi \in C^k([0,r_0],\R ^N
 )$, and $\alpha  \in C^k([0,r_0], \mathrm{End}(\R ^N))$
  with
  $$
  \alpha(0)
  =a \mathrm{Id}
  \;,
  $$
  where $\mathrm{Id}$ is the identity matrix in
  $\mathrm{End}(\R ^N)$.
 If
 $\phi\in C^1((0,r_0], \R ^N )$ is a solution of
\begin{equation}
 \label{23I.3}
 \phi' = \frac \alpha r \phi + \psi
 \;,
\end{equation}
 then:
 \begin{enumerate}
 \item The limit
 \begin{equation}
 \label{23I.1}
 \lim_{r\to 0} r^{-a} \phi
 \end{equation}
 exists.
 \item There exists a solution such that the last limit is
     zero. Fur such solutions $\phi$ extends by continuity
     to a function in $C^{k+1}([0,r_0])$. If moreover
     $\psi=O(r^m)$, respectively $o(r^k)$, then $\phi =
     O(r^{m+1})$, respectively $o(r^{k+1})$. Here by
     $o(r^\infty)$ we mean a function which is $o(r^{k})$
     for all $k$.
 \end{enumerate}
\end{lemma}

\begin{remark}
{\rm
 The fact that $\phi \in C^k((0,r_0)$ is standard, so the only
 issue is at $r=0$. Similarly the case $a=0$ is   standard.
 It is easy to analyze the equation with $a> 0$ using
 similar methods, but the results are more complicated to describe, and
 will not be needed in this work.
 }
\end{remark}

\begin{remark}
 \label{R23I.1}
{\rm
We will be using Lemma~\ref{LODE1} in the following equivalent form:
Suppose that there exist matrices $\alpha_i$ so that $\alpha$
has an expansion
\begin{equation}
 \label{24I.1}
\alpha= a \mathrm{Id} + \alpha_1r+ \ldots + \alpha_k r^k + o_k(r^k)
 \;,
\end{equation}
and suppose that there exist vectors $\psi_i \in \R ^N$
so that $\psi$ has an expansion
\begin{equation}
 \label{24I.2}
\psi= \psi_0 + \psi_1r+ \ldots + \psi_k r^k + o_k(r^k)
 \;.
\end{equation}
Here we write  $f=o_k(r^m)$ if for $0\le i \le k$ we have
$\partial_r^i f = o(r^{m-i})$. Then the limit (\ref{23I.1})
exists. If this limit vanishes, then there exist vectors
$\phi_i \in \R ^N$ so that $\phi$ has an expansion
\begin{equation}
 \label{24I.3}
\phi= \phi_1 r+ \ldots + \phi_k r^k + o_k(r^k)
 \;.
\end{equation} 
}
\end{remark}

\proof  Let us denote by $\langle \cdot , \cdot \rangle$ the
canonical scalar product in $\R ^N$, with $|\phi|^2
=\langle \phi , \phi \rangle$. Set $f:=r^{-2a} |\phi|^2$.  From
(\ref{23I.3}) we have, for some constant $C$,
\begin{eqnarray*}
 r\partial_r(
 \underbrace{r^{-2a} |\phi|^2}_f)
  & = &
  2 r^{-2a} \langle \phi , \underbrace{(\alpha - a \mathrm{Id})}_{\ge -C r }\phi + r \psi  \rangle
\\
 & \ge &
 r\big(-2C  r^{-2a} |\phi|^2 +  r^{-2a } \underbrace{2\langle \phi ,   \psi \rangle}_{\ge -  |\phi|^2 - |\psi|^2}\big)
\\
 & \ge &
 -r\big(2(C+1)f +  r^{-2a }  |\psi|^2\big)
 \;;
\end{eqnarray*}
equivalently
$$
 \partial_r \Big( \underbrace{e^{2(C+1)r} f + \int_{r_0}^r e^{2(C+1)s} s^{-2a }  |\psi(s)|^2\,ds}_{=:h} \Big)\ge 0
 \;.
$$
So the function $h$ defined in the last equation is monotonous,
nondecreasing. 
Monotonicity and positivity of $h$ implies that of existence of
the non-negative limit
$ \lim_{r \to 0 } h(r) $, and  we conclude that $r^{-a}|\phi|$
has a finite limit as $r\to 0$. In particular $|\phi|\le
Cr^{a}$ for some constant $C$.

We rewrite   (\ref{23I.3}) as
$$
\partial_r(r^{-a} \phi) = r^{-a}\left(\psi + (\alpha - a \mathrm{Id}) \phi\right)
 \;.
$$
Integrating, for $0<r_1 <  r\le r_0$ one finds,
\begin{equation}
\label{24I.4xx}
 \frac{\phi(r)}{ r^{a}} = \frac{\phi(r_1)}{r_1^{ a}}   + \int_{r_1}^r s^{-a} \psi(s) \,ds
 + \int_{r_1}^r \underbrace{( \alpha-a \mathrm{Id})  s^{-a} \phi(s)}_{\le C s^{1 }} \, dx
 \;.
\end{equation}
Passing with $r_1$ to zero, using convergence of the  integrals
above in the limit, we find that the limit
$$
 \tilde \phi:= \lim_{r_1 \to 0}  \frac{\phi(r_1)}{r_1^{ a}}
$$
exists. Hence  point \emph{1.} holds, and moreover
\begin{equation}
\label{24I.4x}
 \phi(r) =r^{ a} \tilde \phi  +r^{ a}\int_0^r s^{-a} \psi(s) \,ds
 + r^{ a}\int_0^r  (a \mathrm{Id}- \alpha)  s^{-a} \phi(s) \, dx
 \;,
\end{equation}

\emph{2.} It is standard that  solutions of the homogeneous
equation can be uniquely parameterised by $\tilde \phi$. So,
given any solution of the non-homogeneous equation
(\ref{23I.3}), we can substract from it a solution of the
homogeneous equation with the same value of $\tilde \phi$,
obtaining a solution with $\tilde \phi=0$. It follows from
(\ref{24I.4x}) that we then have
\begin{eqnarray}
\label{24I.4x2}
 \phi(r) &= &  r^{ a}\int_0^r s^{-a} \psi(s) \,ds
 + r^{ a}\int_0^r  (a \mathrm{Id}- \alpha)  s^{-a} \phi(s) \, dx
\\
 & = & \frac r {1-a} \psi_0 + o(r)
 \;.
\nonumber
\end{eqnarray}
Suppose, now, that
\begin{equation}
 \label{24I.3j}
\phi=   \phi_1r+ \ldots + \phi_j r^j + o (r^j)
\end{equation}
holds for some $ 1\le j <k+1$; we have just shown that this
holds with $j=1$. Inserting (\ref{24I.1})-(\ref{24I.2}) and
(\ref{24I.3j}) into (\ref{24I.4x2}) one then finds by
elementary manipulations that (\ref{24I.3j}) holds with $j$
replaced by $j+1$. Lemma~\ref{LODE1} follows now by induction,
using Remark~\ref{R23I.1}.
\hfill $    \Box$

\bigskip

Let $M$ be any smooth compact manifold; in our applications $M$
will be a sphere $S^{n-1}$. By commuting (\ref{23I.3})
 with
differential operators tangential to $M$ one immediately
obtains the following corollary to Remark~\ref{R23I.1}:

\begin{proposition}
 \label{PODE1}
Let $r_0>0$, $k,N\in\N $, $0>a \in \mathbf R$. Suppose
that there exist matrices $\alpha_i\in C^\infty(M,
\mathrm{End}(\R ^N))$ so that $\alpha$ has an expansion
\begin{equation}
 \label{23I.4}
\alpha= a \mathrm{Id} + \alpha_1r+ \ldots + \alpha_k r^k + o_k(r^k)
 \;,
 \end{equation}
and suppose that there exist vectors $\psi_i \in
C^\infty(M,\R ^N)$ so that $\psi$ has an expansion
\begin{equation}
 \label{23I.5}
\psi= \psi_0 + \psi_1 r+ \ldots + \psi_k r^k + o_k(r^k)
 \;.
 \end{equation}
We assume moreover that for any $\ell \in\N $ and for
any smooth differential operator $X$ on $M$ of order $\ell$ the
error terms in (\ref{23I.4}) and in (\ref{23I.5}) satisfy,  for
$0\le i \le k$ and $0\le i+\ell \le k+1$,
\begin{equation}
 \label{23I.6}
 \partial_r^i X(o_k(r^k)) = o(r^{k-i})
 \;.
\end{equation}
Let $\phi\in C^0(M\times (0,r_0], \R ^N )$ be
differentiable in $r$ and satisfy
\begin{equation}
 \label{23I.3zz}
 \phi' = \frac \alpha r \phi + \psi
 \;.
\end{equation}
Then:
 \begin{enumerate}
 \item The limit
 \begin{equation}
 \label{23I.1xz}
 \lim_{r\to 0} r^{-a} \phi
 \end{equation}
 exists.
 \item There exists a solution $\phi\in C^{k+1}(M\times
     (0,r_0], \R ^N )$ of (\ref{23I.3}) such that
     the last limit is zero. Fur such solutions $\phi$ has
     an expansion
$$
\phi=   \phi_1 r+ \ldots   + \phi_{k+1} r^{k+1} + o_{k+1}(r^{k+1})
 \;,
 $$
where $\phi_i\in C^{\infty}(M\times [0,r_0])$, and where
     the error term satisfies (\ref{23I.6}), with $k$ in
     the exponent replaced by $k+1$. If moreover
     $\psi=O(r^m)$, respectively $o(r^k)$, then $\phi =
     O(r^{m+1})$, respectively $o(r^{k+1})$.
 \end{enumerate}
\end{proposition}

\section{Prescribing $\nu_0$}
 \label{A29VIII12.1}

Let $\nu_0$ be the restriction of a smooth space-time function to the future light-cone of a smooth metric $C$.
In this appendix we show how to deform $C$ to achieve   $\overline C_{01}=\nu_0$  without changing $\ol C_{AB}dx^A dx^B$.

Let  $y$ be a coordinate system in which the future light-cone of $C$ takes the  Minkowskian form $y^0=|\vec y|$, and let  $x$ be coordinates as in \eq{19XI.3}. Using the notation
$$
 \underline{C_{\mu\nu}}=C\Big(\frac{\partial}{\partial y^\mu},\frac{\partial}{\partial y^\mu}\Big)
  \;,
  \qquad
  {C_{\mu\nu}}=C\Big(\frac{\partial}{\partial x^\mu},\frac{\partial}{\partial x^\mu}\Big)
  \;,
$$
we have the transformation formulae
\begin{equation}
 \label{22XI.21}
C_{00}\equiv\underline{C_{00}},\quad
C_{01}\equiv-\underline{C_{00}} -\underline{C_{0i}}\Theta^{i},\quad
C_{0A}\equiv-\underline{C_{0i}} r\frac{\partial\Theta^{i}}{\partial x^{A}}
 \;,
\end{equation}
\begin{equation}
 \label{22XI.22}
C_{11}\equiv\underline{C_{00}}+2\underline{C_{0i}}\Theta^{i}
           +\underline{C_{ij}}\Theta^{i}\Theta^{j},\quad
C_{1A}\equiv\underline{C_{0i}} r\frac{\partial\Theta^{i}}{\partial x^{A}}
      +\underline{C_{ji}}r\Theta^{j}
          \frac{\partial\Theta^{i}}{\partial x^{A}}\;,
\end{equation}
\begin{equation}
 \label{22XI.23}
C_{AB}\equiv\underline{C_{ij}}r^{2}\frac{\partial\Theta^{i}}{\partial x^{A}%
}\frac{\partial\Theta^{j}}{\partial x^{B}} \;.
\end{equation}
Conversely, $\underline{C_{\lambda\mu}}=\frac{\partial x^{\alpha}%
}{\partial \regular^{\lambda}}\frac{\partial x^{\beta}}{\partial \regular^{\mu}}%
C_{\alpha\beta}$ gives
\begin{equation}
 \label{20XII.1}
\underline{C_{00}}\equiv C_{00},\quad
\underline{C_{0i}}\equiv
-(C_{00}+C_{01})\Theta^{i}-C_{0A}\frac{\partial x^{A}}{\partial \regular^{i}}
\;,  %
\end{equation}
\begin{equation}
 \label{20XII.2}
\underline{C_{ij}}=(C_{00}+2C_{01}+C_{11})\Theta^{i}\Theta^{j}
+(C_{0A}+C_{1A})
(\Theta^{i}\frac{\partial x^{A}}{\partial \regular^{j}}+\Theta^{j}\frac{\partial
x^{A}}{\partial \regular^{i}})
+C_{AB}\frac{\partial x^{A}}{\partial \regular^{i}}\frac{\partial x^{B}}{\partial \regular^{j}}
 \;.
\end{equation}

As the first step of our argument, we need to write $ \nu_0$ as
\begin{equation}
 \label{12II.1}
 \nu_0 = 1 + r^2 \bar f_0 + \sum_{i=1}^n\bar f_i \regular^i
 \;,
\end{equation}
where $\bar f_0$, respectively $\bar f_i$, are restrictions to
the light-cone of functions $f_0$, respectively $f_i$, which
are smooth on space-time. (It follows from  \eqref{13II.1} that $
\bar f_0=O(r^2)$ when $\kappa=0$, a harmonic gauge and the vacuum Raychaudhuri equation are assumed,
but this will not be needed in what follows.) To prove \eq{12II.1}, let
$f$ be any smooth function on space-time; Taylor expanding $f$
with respect to $\regular^1$  we can write
$$
 f(\regular^0,\regular^1,\ldots,\regular^n) =f(\regular^0,0,\regular^2,\ldots,\regular^n) + f_1(\regular^0,\regular^1,\ldots,\regular^n)\regular^1\;,
$$
where
$$
 f_1=\int_0^1 \frac{\partial f}{\partial \regular^1}(\regular^0,s\regular^1,\regular^2,\ldots,\regular^n)\,ds
  \in C^\infty
 \;.
$$
Similarly
$$
 f(\regular^0,0,\regular^2\ldots,\regular^n) =f(\regular^0,0,0,\regular^3,\ldots,\regular^n) + f_2(\regular^0,\regular^2,\ldots,\regular^n)\regular^2\;,
$$
with
$$
 f_2=\int_0^1 \frac{\partial f}{\partial \regular^2}(\regular^0,0,s\regular^2,\regular^3,\ldots,\regular^n)\,ds
  \in C^\infty
 \;.
$$
Continuing in this way, after $n$ steps the function
\begin{equation}
 \label{10VII0.3}
f-\sum_{i=1}^n f_i \regular^i = f(\regular^0,0,\ldots,0)
\end{equation}
depends only upon $\regular^0$. A final Taylor expansion allows to
rewrite the right-hand-side as $f(0,0,\ldots,0)+(y^0)^m f_0$, where
$f_0$ is a smooth function of $\regular^0$ and  $m$ is the order of
the zero of $f(\regular^0,0,\ldots,0)- f(0,0,\ldots,0)$. Keeping in mind that $y^0|_{C_O}=r$, \eq{12II.1} for $\nu_0=\overline f$
follows.

Let, now, $C_{\mu\nu}$ be given, and consider
$$
 \tC _{\mu\nu} := \Omega^2 C_{\mu\nu} + \delta C_{\mu\nu}
 \;,
$$
where $\Omega=1$ if one wishes to keep $\ol C_{AB}dx^A dx^B$ fixed, or $\Omega$ is a smooth space-time function with prescribed $\ol \Omega$ (e.g. the conformal factor determined in
Section~\ref{ss11II.3}), if one only wishes to prescribe $\ol C_{AB}dx^A dx^B$ up to a conformal factor.

Suppose, momentarily, that the components
$$\underline{\delta
 C_{0 \mu }}
$$
are prescribed smooth functions on space-time, and suppose that
$\delta C$ satisfies
\begin{equation}
 \label{1I.1}
 \delta {C}_{AB} = 0=\overline{\delta {C}_{1 1}}=\overline{\delta  {C}_{1 A}}
 \;.
\end{equation}
The first equality guarantees that the initial data $\tC _{AB}$ defined by $\tC $ coincide with the metric $g_{AB}$
solving the first constraint equation,
 while the last two
guarantee that the cone $\{\regular^0=|\vec \regular|\}$ remains
characteristic for $\tC $.

Then by \eqref{22XI.21}
\begin{equation}
 \label{12II.3}
 \delta C_{00} = \underline {\delta C_{00}}\;,
  \quad
 \delta C_{01} = -\underline {\delta C_{00}}-\underline {\delta C_{0k}}\Theta^k\;,
  \quad
 \delta C_{0A} = -\underline {\delta C_{0k}} \frac{\partial \regular^k}{\partial x^A}
 \;,
\end{equation}
and so all components $\overline{\delta C_{\mu\nu}}$ are known.
We can now find the restrictions to the light-cone of the
missing components $\underline{\delta C_{ij}}$ of $\delta C$
using \eqref{20XII.2}:
\begin{eqnarray}
\nonumber
 \overline
 {\underline{\delta {C}_{ij}}}
 &= &
 \overline
 {(\delta {C}_{00}+2\delta {C}_{01} )\Theta^{i}\Theta^{j}
 + \delta  {C}_{0A}
 (\Theta^{i}\frac{\partial x^{A}}{\partial \regular^{j}}+\Theta^{j}\frac{\partial
 x^{A}}{\partial \regular^{i}})
 }
\\
 \nonumber
  &= &
 \overline
 {-(\underline {\delta C_{00}}+2\underline {\delta C_{0k}}\Theta^k )\Theta^{i}\Theta^{j}
 - \underline {\delta C_{0k}} \frac{\partial \regular^k}{\partial x^A}
 (\Theta^{i}\frac{\partial x^{A}}{\partial \regular^{j}}+\Theta^{j}\frac{\partial
 x^{A}}{\partial \regular^{i}})
 }
\\
 \nonumber
 &= &
 \overline
 {-(\underline {\delta C_{00}}+2\underline {\delta C_{0k}}\Theta^k )\Theta^{i}\Theta^{j}
 - \underline {\delta C_{0k}} \big( (\delta^k_j-\Theta^k \Theta^j)\Theta^i +
 (\delta^k_i-\Theta^k \Theta^i)\Theta^j \big)
 }
\\
 &= &
 \overline
 {- \underline {\delta C_{00}} \Theta^{i}\Theta^{j}
 - \underline {\delta C_{0i}}  \Theta^j
 - \underline {\delta C_{0j}}   \Theta^i
 }
 \;.%
 \label{12II.2}
\end{eqnarray}

Keeping in mind  that $\delta C_{\mu\nu}$ is required to
satisfy \eqref{1I.1}, we  chose the tensor field $\delta C_{\mu\nu}$ now so that in addition to this last equation it holds that
\bel{6III12.1}
 \ol{\tC }_{01}=\nu_0
 \;,
\ee
where $\nu_0$ is the restriction to the light-cone of a smooth function $f$.
Equivalently,
\begin{equation}
 \label{12II.5}
 \delta C_{01} = r^2 f_0 + \sum_{i=1}^n f_i \regular^i +1 -\Omega^2
 \;,
\end{equation}
where $f_0$ and $f_i$ are given by \eqref{12II.1}. As in that last equation we can also write
$$
 \Omega^2 = 1 + r^2 h_0 + \sum_{i=1}^n h_i \regular^i
 \;;
$$
which allows us to rewrite \eqref{12II.5} as
\begin{equation}
 \label{12II.5a}
  \delta C_{01} = r^2 (f_0-h_0) + \sum_{i=1}^n (f_i-h_i) \regular^i
 \;.
\end{equation}
Comparing with \eqref{12II.3}, we see that \eqref{12II.5} will
hold if we choose
$$
 \underline{\delta C_{00}}= r^2 (h_0-f_0)\;, \quad
 \underline{\delta C_{0i}}= t(h_i - f_i) \;,
$$
while, in view of \eqref{12II.2}, \eqref{1I.1} will be
satisfied if $\underline{\delta C_{ij}}$ is further chosen to be
\begin{eqnarray}
 {\underline{\delta {C}_{ij}}}
 &= &
 {(f_0-h_0) \regular^{i}\regular^{j}
 + (f_i-h_i) \regular^j
 + (f_j-\regular_j)\regular^i
 }
 \;.%
 \label{12II.6a}
\end{eqnarray}
The reader might wish to verify by a direct calculation that, with these choices,
\eqref{1I.1} and
\eqref{6III12.1} hold.

The metric $\tC $ will clearly be Lorentzian in a
sufficiently small neighbourhood of the vertex of the cone.

 \bigskip

\noindent{\sc Acknowledgements} I am grateful to  Yvonne Choquet-Bruhat and Jos\'e Mar\'ia Martin-Garc\'ia for many useful discussions,
and collaboration on previous attempts to solve the problems addressed in this work. Supported in part by Narodowe   Centrum Nauki under the grant DEC-2011/03/B/ST/02625.

\bibliographystyle{amsplain}
\bibliography{../references/hip_bib,%
../references/reffile,%
../references/newbiblio,%
../references/newbiblio2,%
../references/chrusciel,%
../references/bibl,%
../references/howard,%
../references/bartnik,%
../references/myGR,%
../references/newbib,%
../references/Energy,%
../references/dp-BAMS,%
../references/prop2,%
../references/besse2,%
../references/netbiblio,%
../references/PDE}

\end{document}